\documentclass [11pt]{article}
\usepackage{graphicx}
\usepackage{subfigure}

\begin{document}

\begin{center}
{\LARGE \bf On final states of 2D decaying turbulence}
\end{center}

\begin{center}
Z. Yin\footnote{Electronic address: yinzh@lsec.cc.ac.cn}
\end{center}

\begin{center}
Fluid Dynamics Laboratory, Applied Physics Department, Eindhoven
University of Technology, P.O.Box 513, 5600 MB, Eindhoven, The
Netherlands
\end{center}

\begin{center}
LSEC, Institute of Computational Mathematics, Chinese Academy of
Sciences, P.O.Box 2719, Beijing 100080 P.R.China
\end{center}

Numerical and analytical studies of ``final states'' of
two-dimensional (2D) decaying turbulence are reported. The first
part of this work is trying to give a definition for final states
of 2D decaying turbulence. Although the functional relation of
$\omega - \psi $ is frequently used as the characterization of
those ``final states,'' it is just a sufficient but not necessary
condition so it is not proper to be used as the definition. It is
found the way through the value of the effective area $S$ covered
by the scatter $\omega - \psi $ plot, which is initially suggested
by Read [1], is more general, and more suitable for the
definition. Based on this concept, we gave out a definition that
can cover all existing results in late states of decaying 2D
flows, including some weird double-valued $\omega - \psi $ scatter
plots that can not be explained before. The rest part of the paper
is trying to further investigate 2D decaying turbulence with the
assistance of our new definition. Some new numerical results,
which lead to ``bar'' final states and further verify the
predictive ability of statistical mechanics [2], are reported. It
is realized that some simulations with narrow-band energy spectral
initial conditions, which can be called ``turbulence'' doubtfully,
lead to some final states that can not be very well explained by
the statistical theory (in the meanwhile, they are still in the
scope of our new definition of the ``final state''). For those
simulations with initial conditions of broadband energy spectra
that lead to the famous dipole, we give out a mathematical
re-interpreting for the so-called sin-hyperbolic (``sinh'')
$\omega - \psi $ scatter plot in final states. We suggest the term
``sinh'' here should be replaced by ``sinh-like.'' The
corresponding physical meaning of this re-interpreting will also
be discussed.

\section*{I. INTRODUCTION}

It is an interesting topic to study final states of
two-dimensional decaying turbulence. Recent publications [2]
(hereafter YMC) and [3] show that numerical simulations starting
from different initial conditions (depending on the size of
patches in the vorticity field) can lead to different final
states. The statistical mechanics in two-dimensional turbulence,
known as the ``point'' theory and the ``patch'' theory, shows a
great predictive power in this kind of simulations.

The ``point'' theory is concerned with a mean-field treatment of
ideal line vortices, which had been given thirty years ago [4, 5].
The system is Hamiltonian with a finite phase space, applied by
Boltzmann statistics to its dynamics initially by Onsager [6]. It
was further developed by several groups [7-23]. In these
investigations, it is surprise to see that the ideal Euler
mean-field predictions fit the Navier-Stokes results.

The ``patch'' theory started from the late 1980s [24-29]. In the
``patch'' theory, the delta-functions that are used to discretize
the vorticity field in the ``point'' theory are replaced by
finite-area, mutually exclusive ``patches'' of vorticity. The
Lynden-Bell statistics [30] is applied to this theory.

The predictive abilities of these two theories are tested after
defining the related entropy of them and developing the precise
formulas suited for calculations, the details of which are given
in {\it Spring Notes} by D.C. Montgomery (private communication;
see also a modified version of it in chapter 2 of [31]). Several
kinds of solutions in the double periodical domain from the
statistical mechanics are considered. It is found that the
traditional ``dipole'' or the one-dimensional ``bar'' solution
will have the largest entropy under different conditions: for the
vorticity field with large patch vortices, the ``bar'' solution is
the maximum entropy state, and the ``dipole'' will dominate the
final state if only point vortices or small patches exist
initially. The prediction is validated by most of our direct
numerical simulations, which are well represented by 13
simulations listed in [31]. These simulations, which make use of
Fourier pseudo-spectral methods, have considerable high
resolutions ($512^2)$ and have been run long enough (100 - 1500
turnover times) to make sure to reach final states. Those runs
tested theoretical results through all kinds of aspects: from
maximum entropy states to local maximum states. We even found a
couple of results that lead to unclassified states, which are
excluded by the most general case of the ``patch'' theory (see
section II and III for some of them).

However, comparing to considerable efforts to investigate the
``final state'' of 2D turbulence, few efforts have been devoted to
define it. There exist some characters when the term of ``final
states'' in 2D flows is referred to, but they are rather blurred
and it is very easy to find some negative examples for them as the
investigation in this field goes on.

For example, sometimes it is thought that the flow field has
reached the ``final state'' if the pattern of the flow remains
unchanged for a certain long time. However, in our sinh-Poisson
quadrupole to ``dipole'' simulation (Figs. 7 in [3]), the
quadrupole in the vorticity field lasts so long (from $t \cong 0$
to $t \cong 150)$ that people might think the ``final state'' has
already been reached, while the continued calculation showed that
it is just a local maximum state.

There are still some other characters except the example above (we
will discuss another one in detail later in section IIA). Those
characters are important phenomena of ``final states.'' In the
meanwhile, they can be very misleading. Thus it is necessary to
give out a definition which can includes all exiting cases of
``final states.''

This definition will play an important role when a long direct
numerical simulation of decaying turbulence is carried out. It can
be used to decide when the code should be stopped. This is not a
trivial decision since some of the results (including the ``bar''
final state) in YMC are obtained partly because of the continuing
calculation after former researchers stopped [32] (of course,
those runs are mainly stimulated by theoretical results of the
statistical-mechanics). In our former investigation, the
pseudo-spectral code has been continued as long as possible to
make sure that final states have been reached. The calculation can
be significantly extended without a good definition of the ``final
state.'' On the other hand, because of the existing bottleneck in
parallelization of the 2D pseudo-spectral code [3], the extended
calculation might mean several extra weeks (the wall clock time)
for runs with the resolution of $512^2$. The future investigation
in this field may involve in simulations with the resolutions of
$1024^2$ or higher. For those high-resolution simulations, even
with the fastest parallel spectral Fourier 2D code we have so far
(using the combined technique of domain decomposition and task
distribution [33]), it would be a nightmare for the investigation
if no clue is used to judge the final state and a lot of extended
computation is needed to make sure that there is no interesting
new phenomenon any more. In section II, we will try to give out a
definition of ``final states'' for those kinds of purposes.

In section III, with the assistance of the new definition of
``final states,'' (on the other hand, in order to validate the
definition), some numerical simulations are carried out. This part
of the work can be also treated as the continued investigation of
YMC. Former numerical results in YMC leading to ``bar'' final
states are confined by an initial symmetric quadrupole with a
certain mount of noise added to break the symmetry and accelerate
the process of computations. Without further proofs, it can be
argued that the ``bar'' is reached only because of the existing
symmetry in the main structure. Hence the generality of this kind
of solutions might be endangered. In section III, some other
techniques are adopted to break the symmetry, and again they end
up with ``bar'' states. These simulations also confirm our former
statement (in a much more direct sense): ``it is the size of the
`patch' that plays a key role in the process towards the `bar'
final state.''

In section IIIB, another totally different initial condition
leading to the ``bar'' will be reported. Although the statistical
mechanics can predict the flow pattern for the final state of that
simulation (the patch size in the initial vorticity field is big,
and thus the ``bar'' is expected [2]), it has some difficulties
when explaining the $\omega - \psi $ scatter plot at the late
time. Here, it should be noticed that that ``final state'' is
still in the scope of our new definition in section IIB.

Section IV is concerned with an influential simulation in this
field [34-36]. The question was raised when we tried to repeat the
pioneers' work - changing the parameters of the sin-hyperbolic
(sinh) function to make a best fit for the scatter $\omega - \psi
$ plot of the famous ``dipole.'' This fitting process gave us the
hallucination that the sum of several sinh functions is just
another sinh function. Of course, this is wrong, and the fitting
error can be reduced dramatically by using the combination of
several instead of just one sinh function. The corresponding
physical meaning of this fitting process will also be discussed.
The main content of this section is more concerned with the
conception of the statistical mechanics than presenting new
results: it is suggested to replace the term ``sinh'' in existing
literatures of this field by the term ``sinh-like'' to represent
the ``point'' theory and numerical results better.

There are mainly two threads when we are writing this paper: the
first one is the new definition of ``final states,'' another one
is that this paper is the extended work of YMC.

\section*{II. DEFINING THE ``FINAL STATE'' IN 2D TURBULENCE}

\subsection*{A. Starting from a sufficient but not necessary condition of ``final
states''}

In 2D flow field, if we denote the vorticity as $\omega $, and the
stream function $\psi $, the Navier-Stokes equation can be written
as

\begin{equation}
\label{eq1} \frac{\partial \omega }{\partial t} + (\frac{\partial
\omega }{\partial x}\frac{\partial \psi }{\partial y} -
\frac{\partial \omega }{\partial y}\frac{\partial \psi }{\partial
x}) = \nu \nabla ^2\omega
\end{equation}

\noindent and

\begin{equation}
\label{eq2} \omega = - \nabla ^2\psi ,
\end{equation}

\noindent with $\nu$ the kinematic viscosity of the fluid.

A shorthand notation for the nonlinear contribution to Eq.
(\ref{eq1}) is the Jacobian $J(\omega ,\psi )$

\begin{equation}
\label{eq3} J(\omega ,\psi ) = \frac{\partial \omega }{\partial
x}\frac{\partial \psi }{\partial y} - \frac{\partial \omega
}{\partial y}\frac{\partial \psi }{\partial x}.
\end{equation}

For inviscid flows, Eq. (\ref{eq1}) reduces to the Euler equation

\begin{equation}
\label{eq4} \frac{D\omega }{Dt} = \frac{\partial \omega }{\partial
t} + J(\omega ,\psi ) = 0,
\end{equation}

\noindent which states that the vorticity of a fluid element is
conserved for inviscid flows (note that ${D\omega}/{Dt}$
represents the material derivative).

It can be seen that if there exists a functional relation $\omega
= f(\psi )$, then

\begin{equation}
\label{eq5} J(\omega ,\psi ) = \frac{\partial f(\psi )}{\partial
x}\frac{\partial \psi }{\partial y} - \frac{\partial f(\psi
)}{\partial y}\frac{\partial \psi }{\partial x} = (\frac{\partial
f}{\partial \psi }\frac{\partial \psi }{\partial x})\frac{\partial
\psi }{\partial y} - (\frac{\partial f}{\partial \psi
}\frac{\partial \psi }{\partial y})\frac{\partial \psi }{\partial
x} = 0.
\end{equation}

This means that the experimental or numerical observation of the
functional relationship $\omega = f(\psi )$ indicates the presence
of a stationary state of inviscid flow. Usually, this observation
is treated as an indication of the presence of a nearly steady
state of high Reynolds number flows, $i.e.$, the case when $\nu
\to 0$.

{\it Although the }$\omega - \psi $ {\it  functional relation is
an important tool in the characterization of so-called ``final
states'' of decaying 2D turbulence, we should notice that it is a
sufficient but not necessary condition of near-equilibrium states
of high Reynolds number flows.}

In Figs. 1, we see a good example of the functional relationship
between $\omega $ and $\psi $ - the famous Lamb dipole (for
further details see [37, 38]). It has the linear relation of
$\omega - \psi $ (for $ - 0.5 < \psi < 0.5)$. Another example is
the sinh relationship [34-36], of which we will give a
re-interpreting as the ``sinh-like'' relation in section IV.

In Figs. 2, we see a double-valued structure of the $\omega - \psi
$ plot [39] and the associated $\omega $ and $\psi $ contour
plots. They come from one simulation in YMC - Figs. 18 and 19,
which cannot be explained by the existing statistical-mechanical
theories. By comparing the $\omega - \psi $ scatter plot with the
contour plots of the vorticity and the stream function, we can
conclude that the larger negative vortex, which is indicated by
the solid arrow, corresponds to the longer negative branch of the
$\omega - \psi $ scatter plot. In the meanwhile, the shorter
(negative) branch of the $\omega - \psi $ scatter plot represents
the smaller negative vortex (see the dashed arrow).

The vorticity field shown in Fig. 2(a) represents a stationary
solution of the Euler equation ${D\omega}/{Dt} = 0$. This can be
demonstrated numerically: what we did is using the pseudo-spectral
Fourier code of the Navier-Stokes equation and setting $\nu = 0$
(this is the handiest way to test it in our case); the exact
vorticity field of Fig. 2(a) (without any noise) is used as the
initial condition for a simulation with the resolution of $512^2$.
The simulation lasts for $3\times 10^5$ time steps (the time step
is $1/2000$), which is about 400 turnover times if we set $\nu =
1/5000$. However, the vorticity field has not been changed a bit
from $t = 0$ to $t = 150$. The continuing running of the code is
more a test of the accuracy of the pseudo-spectral method than
anything else.

This double-valued structure cannot be explained by the
statistical-mechanical theory for Euler flows, even if the most
general formulation of the ``patch'' theory

\begin{equation}
\label{eq6} \nabla ^2\psi = - \omega = - \sum\limits_{j = 1}^q
{\frac{M}{\Delta }K_j \frac{e^{\alpha _j - \beta \psi K_j
}}{\sum\limits_{l = 0}^q {e^{\alpha _l - \beta \psi K_l }} }}
\end{equation}

\noindent is taken into consideration.

At this point, we should admit that although the application of a
statistical-mechanical approach to predict the quasi-stationary
final state of inviscid flows appears to be very powerful to
investigate freely evolving 2D turbulent flows, it still has some
limitations, which cannot be easily understood.

The flow is not described by any functional relation between
$\omega $ and $\psi $, but it manages to go to one specific kind
of equilibrium states. This is partly due to the fact that this
simulation starts from a condition with narrow-band energy spectra
- more specifically, most of the energy is concentrated in few low
wave numbers. Because of inverse energy cascade phenomena in 2D
turbulence (the energy is mainly transferred to lower wave
numbers), the state of broad-band energy spectra, where the
statistical mechanics might take effect, can never be reached.

Anyway, the functional relation of $\omega - \psi $ is not enough
to define the ``final state,'' we need something else to be used
as the definition. And this will be the main task in the following
subsection.

\subsection*{B. A definition covering all existing possibilities}

To estimate the quantitative flux across the region, people
normally use a diagnostic technique first described by Read {\it
et al. }[1]. They showed that the net flux of vorticity out of a
closed loop in the physical space is equal to the effective area
enclosed by the corresponding circuit in the $\omega - \psi $
space. However, the usage of the effective area so far is limited
to indicate how far away from the state of the $\omega - \psi $
functional relation the flow field is [40, 41]. In the following,
we will show that it is in fact a more powerful and more general
judgment for equilibrium states of 2D turbulence than the simple
$\omega - \psi $ functional relation.

For the integrality of this paper, this judgment will be re-stated
as follows:

\begin{itemize}

\item

 At a certain stage of a numerical simulation of 2D decaying
turbulence, draw a closed circuit in the contour plot of $\psi $
that can represent the whole flow field, find enough points on
this circuit and mark them in order. In Fig. 3(a), we only use 5
points for convenience.

\item

Find the corresponding points in the scatter plot of $\omega -
\psi $ at the same time of the simulation. There will be two kinds
of possibilities:

\begin{enumerate}
\item

 Those points form a simple circuit, the area of which is
equal to the effective area $S$ (Fig. 3(b)).

\item
Those points form a region that is reentrant. The effective
area $S$ in this case is equal to the sum of anticlockwise areas
minus the sum of clockwise areas. For example, in Fig. 3(c), the
effective area $S$ is

\begin{equation}
\label{eq7} S = S_{anticlockwise} - S_{clockwise} = S_1 - S_2 .
\end{equation}

(In Fig. 3(c), we only draw one clockwise and one anticlockwise
region for convenience. In a real problem, it may have several
clockwise and anticlockwise areas respectively.)

\end{enumerate}

\item

The absolute value of $S$ indicates how far away the flow field is
from the equilibrium state. The larger the absolute value of $S$,
the further away the flow field from the final state. On the other
hand, if the absolute value of $S$ is very small ($S \cong 0)$, it
means that the equilibrium state has been reached.

\item

The condition of $S \cong 0$ is only enough to judge the
equilibrium state of 2D turbulence. To define the ``final state,''
it is necessary to remove those local maximum states (see for
examples in section IIIB2 of YMC).

\end{itemize}

To sum up, the ``final state'' of 2D turbulence can be defined as:

{\it The flow field of 2D turbulence has reached the ``final
state'' if the following two conditions are true:}
\begin{itemize}
\item {\it The effective area }$S \cong 0;$
\item {\it It is not a
local maximum state.}
\end{itemize}

Note that the ``real'' final state of the flow field is the zero
vorticity state ($\omega \equiv 0)$ when the decaying process
really stops. But those cases are not interesting to us. The
``final state'' that we are talking about is actually a stage when
the continuing numerical simulation will not lead to any new
interesting phenomenon.

Here, the condition of $S \cong 0$ cannot be strictly defined. How
small the absolute value of $S$ should be depends on different
specific conditions. According to our experience, it is more
useful to look into the scatter plot of $\omega - \psi $ itself
than to give out any specific value.

In the following, we will use the new definition to analyze
different situations:

\begin{itemize}
\item
 For those results with the functional relation of $\omega -
\psi $ such as the lamb dipole or the ``sinh-like'' relation
discussed in the previous subsection, it is obvious that $S \cong
0$ - the ``final state'' has been reached.

\item For those results with multi-valued structures in $\omega -
\psi $ plots, there are three kinds of situations:
\begin{enumerate}
\item If that multi-valued structure does not close any area ($S
\cong 0)$, such as Fig. 2(c), it can be also said that the final
state has been reached. (Actually Fig. 2(c) is not a good example:
if we put some noise onto the weird quadrupole, and use it as the
initial condition of a DNS run, we will end up with a normal sense
of the $\omega - \psi $ functional relation. So actually this
weird quadrupole is just a local maximum state. However, we have
to make the definition be able to sort out similar situations with
maximum entropies, which might appear in the future research.)

\item If that multi-valued structure does close some areas, but
they are arranged clockwise and anticlockwise, and can cancel each
other (again, $S \cong 0)$. We can also say that the ``final
state'' has been reached. (We will show an example in section
IIIB.)

\item If that multi-valued structure closes some areas, but
clockwise and anticlockwise parts of them cannot cancel each
other, then the final state has not been reached, and continuing
calculations are needed. An extreme example is a state when the
scatter points of $\omega - \psi $ are distributed across the
whole plot, which normally happens when the simulation was started
using some random initial condition.

With this judgment, we can easily sort out those non-equilibrium
states such as the traveling wave in the patch quadrupole to the
``bar'' simulation (Figs. 4).
\end{enumerate}
\end{itemize}

\section*{III. FURTHER INVESTIGATIONS ON THE ``BAR'' FINAL STATE}

In YMC, we have investigated the emergence of the ``bar'' final
state by considering an antisymmetric basic flow (the quadrupole
solution) with a considerable amount of noise added to it in order
to break the symmetry of the basic flow. We decided to devote some
more efforts to find other initial conditions leading to the
``bar'' quasi-stationary final state, and it will be seen shortly
that the appearance of the ``bar'' final state is not that
accidental as might erroneously be concluded from the previous set
of simulations.

The simulations are finished by the dynamical pseudo-spectral-code
of the 2D NS equation, using a resolution of $512\times 512$
Fourier modes. The time step in all simulations is fixed at
$1/2000$ and determined by the CFL condition. The initial energy,
using the normalization of

\[
E = \frac{1}{2}\frac{1}{(2\pi )^2}\int\!\!\!\int {\omega \psi
{\kern 1pt} dx{\kern 1pt} dy}
\]

\noindent is 0.5. There is no hyperviscosity or small-scale
smoothing of any kind in our simulations.

\subsection*{A. Simulations by shrinking the size of ``patches'' in the initial field}

The first idea here is to start from the same quadrupolar
``patch'' initial condition as YMC, but distort the initial
condition slightly in the following way: we shrink the patch size
with a small amount and reposition the patches slightly (see, for
example, the plots in the first row of Figs. 5). The symmetry of
the basic flow has been broken already, and there is no need to
add any noise to it (unlike what we did in Fig. 7(b) of YMC). The
Reynolds number is fixed at $1/\nu = 8000$. As can be seen from
Figs. 5, we have performed two simulations with two different
distorted quadrupolar initial conditions, with the patch size
reduced by a factor of $7/8\times 7/8 = 49/64$ compared to the
patch size of the original quadrupole initial condition (see Fig.
7(a) in YMC), and both runs clearly reveal the emergence of the
quasi-stationary ``bar'' final state. A similar set of simulations
has been carried out, but now with the patch size even further
reduced. In Figs. 6, we have shown the vorticity contour plots of
runs with the patch size reduced by a factor of $3/4 \times 3/4 =
9/16$, and it is clear that no ``bar'' final state is found in
this case.

We may recall that in Figs. 4 of YMC, the E-S plot predicts that
for doubly-periodical domain, large ``patch'' vortices lead to the
``bar'' quasi-stationary final state and small ``patch'' vortices
lead to the ``dipole'' final state. In YMC, we only test two
extreme cases of theoretical results -- the initial quadrupole
solutions with the largest patches (Figs. 7-10 in YMC) and the
smallest patches - ``point'' (Figs. 14-16 in YMC). There is no
intermediate simulation that can make the logic more complete. The
four simulations in Figs. 5,6 provide a much more direct proof for
our theoretical results of the statistical-mechanics. We can
predict that if the size of the patch is shrunk further (by a
factor even smaller than $9/16$), the numerical simulation will
lead to the dipole final state.

One question that might be raised concerns the direction of the
``bar'' final state. As can be observed in Figs. 5, it can happen
in the horizontal or vertical direction (and should occur with
equal probability due to their symmetrical equivalence). However,
the ``bar'' final state (with $2\pi$ periodicity perpendicular to
the flow direction) has never been observed in any other direction
due to lack of periodicity of such a solution. Note that a
solution with periodicity less than $2\pi$ perpendicular to the
flow direction enables a flow rotated with respect to the x and y
direction (e.g., a direction of $45^ \circ $ is needed for a bar
solution with a $\sqrt 2 \pi $- periodicity).

Due to the inverse energy cascade phenomenon in 2D turbulence,
most of the energy will be concentrated on the lowest modes of
wavenumber at final states - there are four modes altogether:

\[
\vec {k} = (k_x ,k_y ) = (1,0),( - 1,0),(0,1){\kern 1pt}
\,or\,{\kern 1pt} (0, - 1).
\]

Different combinations of these four modes dominate flow patterns
at final states. Here are three kinds of possibilities altogether:
\begin{itemize}
\item If energies are more or less equally distributed in these
four modes, we will get the dipole (see the last vorticity contour
plot in Figs. 5 of YMC);

\item If energies are concentrated on either (1,0), (-1,0) or
(0,1), (0, -1), then we will get ``bar'' final states (like Figs.
5);

\item If energies are distributed on those modes unequally, then
we will get a final state between ``dipole'' and ``bar'' (like
Figs. 6, especially Figs. 6(b)).
\end{itemize}

\subsection*{B. The ``slanting bar'' to ``bar'' simulation}

We have conducted another set of simulations with initial
conditions that was found to lead to the ``bar'' final state. The
initial condition is based on the slanting bar solution already
referred to in the previous subsection ($\sqrt 2 \pi
$-periodicity), where a certain amount of noise is added to it to
break the symmetry. We let the flow evolve and as shown in Figs.
7, the ``bar'' final solution is obtained eventually. The Reynolds
number in this simulation is fixed at $1/\nu = 8000$. The initial
Taylor-scale Reynolds number

\[
R_\lambda \approx \sqrt {\frac{10}{3}} \frac{E}{v\sqrt \Omega }
\approx 4002.
\]

It increases to 8035 at the end of the simulation. Attention
should be drawn to the $\omega - \psi $ scatter plot obtained at
the end of the simulation. It is similar to the scatter plot
obtained for the ``bar'' solution in Fig. 9 of YMC, but some
subtle differences can be observed. When considered in more
details, the scatter plot is in fact a double-valued structure.
However, this double-valued structure is different from the open
line in Fig. 2(c). It is a structure that actually encloses some
areas. We maybe recall the discussion in the previous section that
if those areas cannot cancel each other, namely $S \ne 0$, it may
mean that the ``final state'' has not been reached and the
continuing calculation is needed.

However, with a close examine of the $\omega - \psi $ scatter
plot, we will find that the region covered by scatter points is
actually reentrant, and thus $S \approx 0$:

\begin{quotation}
 This procedure can be done by drawing a line from the
bottom to the top of the contour plot of $\psi $ (see the arrow
line at the last figure in the right column of Figs. 7). It is not
necessary to draw a loop (like what we did in Fig. 3(a)) in this
case, because we are dealing with the doubly-periodical condition,
and a straight line that connects two opposite boundaries is
already enough to make a loop. At the late stage of this
simulation, the flow field is essentially one-dimensional. It is
possible to gather all the information by studying this straight
line. As indicated in Fig. 8, the corresponding points in $\omega
- \psi $ plot form a clockwise region and an anticlockwise region,
which can almost cancel each other (see Eq. (\ref{eq7})). So in
fact, the absolute value of the effective area $S$ is very small -
the ``final state'' has been reached.
\end{quotation}

This simulation illustrates one special case that our new ``final
state'' definition covers (see the second last paragraph in
section IIB).

Again, the statistical mechanics cannot explain this double-valued
structure due to the less ``turbulent'' initial condition.

\section*{IV. A RE-INTERPRETING FOR THE $\omega - \psi $ PLOT OF THE FAMOUS DIPOLE}

About a decade ago, Matthaeus {\it et al. }[34-36] reported the
first long time simulation which led to the ``dipole'' final
state. They used the sinh function to fit the scatter $\omega -
\psi $ plot at the late time [36] for the first time, and they
found a close fitting function by changing the values of $\alpha $
and $\beta$ in

\begin{equation}
\label{eq8} \omega = \alpha \sinh (\beta \psi ).
\end{equation}

However, we will argue that it is not enough to use only one
single sinh function, more sinh functions should be adopted to fit
the scatter plot. The first subsection will tell people a fake
theorem, which is the reason why the pioneer works are restricted
themselves to one sinh function. The second subsection will be our
re-interpreting of the $\omega - \psi $ plot.

\subsection*{A. Starting from a fake theorem}

The fake theorem states as follows:

$\forall {\kern 1pt} \alpha _1 ,\alpha _2 ,\beta _1 ,\beta _2 \in
${\it real numbers, there are two real numbers }$\alpha _0
\,and\,\beta _0 ${\it , which make the following equation true:}

\begin{equation}
\label{eq9} \alpha _1 \sinh (\beta _1 x) + \alpha _2 \sinh (\beta
_2 x) = \alpha _0 \sinh (\beta _0 x).
\end{equation}

If the theorem above is true, we can easily get a more general
conclusion from it:

$\forall {\kern 1pt} \alpha _i ,\beta _i \,(i = 1,2,3,\ldots n)
\in ${\it real numbers, there are two real numbers }$\alpha _0
\,and\,\beta _0 ${\it , which make the following equation true:}

\begin{equation}
\label{eq10} \sum\limits_{i = 1}^n \alpha _i \sinh (\beta _i x) =
\alpha _0 \sinh (\beta _0 x).
\end{equation}

Eqs. (\ref{eq9}) and (\ref{eq10}) are not reasonable at the first
thought, but Figs. 9, which are the plots of the following
functions, seem to cater for them:

\begin{equation}
\label{eq11} y = 0.00008\sinh (2x),
\end{equation}

\begin{equation}
\label{eq12} y = 0.05\sum\limits_{i = 1}^{100} {\sin (i) * \sinh
(\frac{x}{i})} ,
\end{equation}

\noindent and

\begin{equation}
\label{eq13} y = 0.0001\sum\limits_{i = 1}^{100} {\sin (i) * \sinh
(2^{\frac{1}{i}}x)} .
\end{equation}

Figs. 9(b, c) look so much like a sinh function that they give us
the feeling that we can always find a good combination of the
parameters $\alpha _0 ,\beta _0 $, which makes the simple sinh
function ($\alpha _0 \sinh (\beta _0 x))$ fit any of those figures
(as what Eq. (\ref{eq10}) indicated).

We have drawn the plot of $y = \sum\limits_{i = 1}^n \alpha _i
\sinh (\beta _i x)$ with other different values of $n$, $\alpha _i
$, and $\beta _i $, it always looks like a single sinh function.

Of course, Eqs. (\ref{eq9}) and (\ref{eq10}) are not true. They
can be easily disproved after comparing the first few coefficients
of the Taylor series in both sides of the equal mark.

\subsection*{B. ``Sinh'' or ``sinh-like''?}

Partly misled by the same mistake we made in the fake theorem
above, the former researchers restricted their efforts to fit the
scatter $\omega - \psi $ plot of ``dipole'' by using only one sinh
function, which corresponds to two kinds of particles in the
``point'' theory. In this subsection, it will be seen that we can
get a better fitting function if more particles are adopted.

Before making any comparison of two fitting methods, it is
necessary to define an indicator of the fitting degree. There are
many choices to do that, and the correlation factor is adopted
here

\[
R^2 = 1 - \frac{\sum\limits_i {(\omega _i - f(\psi _i ))^2}
}{\sum\limits_i {(\omega _i - \bar {\omega }_i )^2} },
\]

\noindent where $f$ is the function used to fit the curve, and
$\bar {\omega }_i $ is the space average of $\omega $. The larger
$R^2$ is, the better fitting function $f$ we will obtain (the
maximum value of $R^2$ is 1).

The scatter $\omega - \psi $ plot to fit is from Fig. 6 of YMC,
which is essentially a reproduction of the result of Matthaeus
{\it et al. }[34-36]. As it can be seen in Fig. 10(a), if only two
kinds of particles are adopted (the fitting function is limited to
one sinh function), there will be some part of the fitting line
running away from the scatter plot (indicated by the circle). With
four kinds of particles (Fig. 10(b)), a much better fit is
obtained. Of course, the more particles are adopted, the better
fitting function we will obtain.

Note that in Fig. 10(b), two more unequal strength particles -
$e^{1.07x}$ and $e^{ - 1.27x}$ instead of two equal ones (or,
another sinh function) are introduced. This is because the
asymmetry exists in the $\omega - \psi $ plot; the unequal
strength particles make the fitting better than the sinh
functions. Pointin {\it et al.} use the similar technique in their
research before [9], but they also only use two kinds of
particles.

For the physical meaning of this fitting process, it is the same
problem as the following question:

{\it How many kinds of vortices are enough to represent the real
vorticity field? }

Of course, the more kinds of vortices are adopted, the better.

The discussion above gives out the explanation why sinh-like plots
(we know they are not necessary sinh functions anymore) are
observed so often in the late time of numerical simulations of 2D
decaying turbulence:

\begin{quotation}
In the patch theory, there are millions of $\omega - \psi $
relations, the general form of which is illustrated in Eq.
(\ref{eq6}). If sizes of all the patches are shrunk to point, we
will always obtain sinh-like plots. Those sinh-like plots do not
follow the same formula, but they look the same.
\end{quotation}

\section*{V. CONCLUSIONS}

In this paper, we try to give out a definition of the ``final
state'' of 2D decaying turbulence. Our new definition, which makes
use of the effective area $S$ in the $\omega - \psi $ space, is
more general than the ordinary functional relation, and can cover
all existing results.

We found some new DNS results that can further confirm the
predictive power of the statistical mechanics. It is realized that
existing numerical results that verify the ``patch'' theory are
those runs starting from narrow-band low wavenumber initial
conditions. On the other hand, such less ``turbulent'' initial
conditions tend to lead to ``patch'' favor results or some weird
states that no existing statistical mechanics can explain.

We also found the term ``sinh,'' which is frequently used in
existing literatures of 2D turbulence research, should be replaced
by ``sinh-like.'' It is actually very natural to do that, because
the more kinds of particles are used to represent the vorticity
field, the more accurate results will be obtained.

Finally, for the integrality of investigations about the ``patch''
and ``point'' theory in the statistical mechanics, we should not
only connect the statistical theory with numerical simulations (in
YMC and this paper), but also connect the theory with experiments:
\begin{itemize}
\item

 For the ``patch'' theory, so far as to our knowledge, there
is no experiment setup that can generate flat vortices, which are
illustrated in our numerical simulations. It would be interesting
to see such kind of experiments in the future.

\item

 For the ``point'' theory, it is more easily to generate random point
vortices in the laboratory, but our numerical code and theory
results are only dealing with double periodical boundary condition
for the time being. It is easier to archive high resolutions and
perform high Reynolds number simulations by doing this, but it is
also difficult to find any experimental comparison. The next step
of this research will try to connect the theory with more complex
boundaries - for example, the no-slip boundary, which can easily
find some laboratory proofs.
\end{itemize}

\section*{ACKNOWLEDGEMENTS}

Numerical simulations in this paper are finished on the SGI Origin
3800 at SARA Supercomputing Center in Amsterdam and the Legend
DeepComp 1800 at LSEC, Academy of Mathematics and System Science
in Beijing.

I would like to thank Prof. G.J.F van Heijst, Prof. D.C.
Montgomery and Dr. H.J.H. Clercx who have contributed to this
paper directly or indirectly (make corrections in the
corresponding part of my thesis [31]). I also thank Prof. E. van
Groesen, Prof. T.J. Schep, Prof. Annick Pouquet, and Prof. F.W.
Sluijter for useful discussions.

I also want to thank D. van der Woude, D. Molenaar, Prof. W. van
de Water, W. Kramer, and H. Castelijns, who took part in the
``sinh'' or ``sinh-like'' discussion. I thank their efforts to
disprove my fake theorem in section IVA, special regard to some of
them who tried to prove the fake theorem.

\begin{figure*}[!htbp]
\begin{minipage}[c]{.42 \linewidth}
\scalebox{1}[1]{\includegraphics[width=\linewidth]{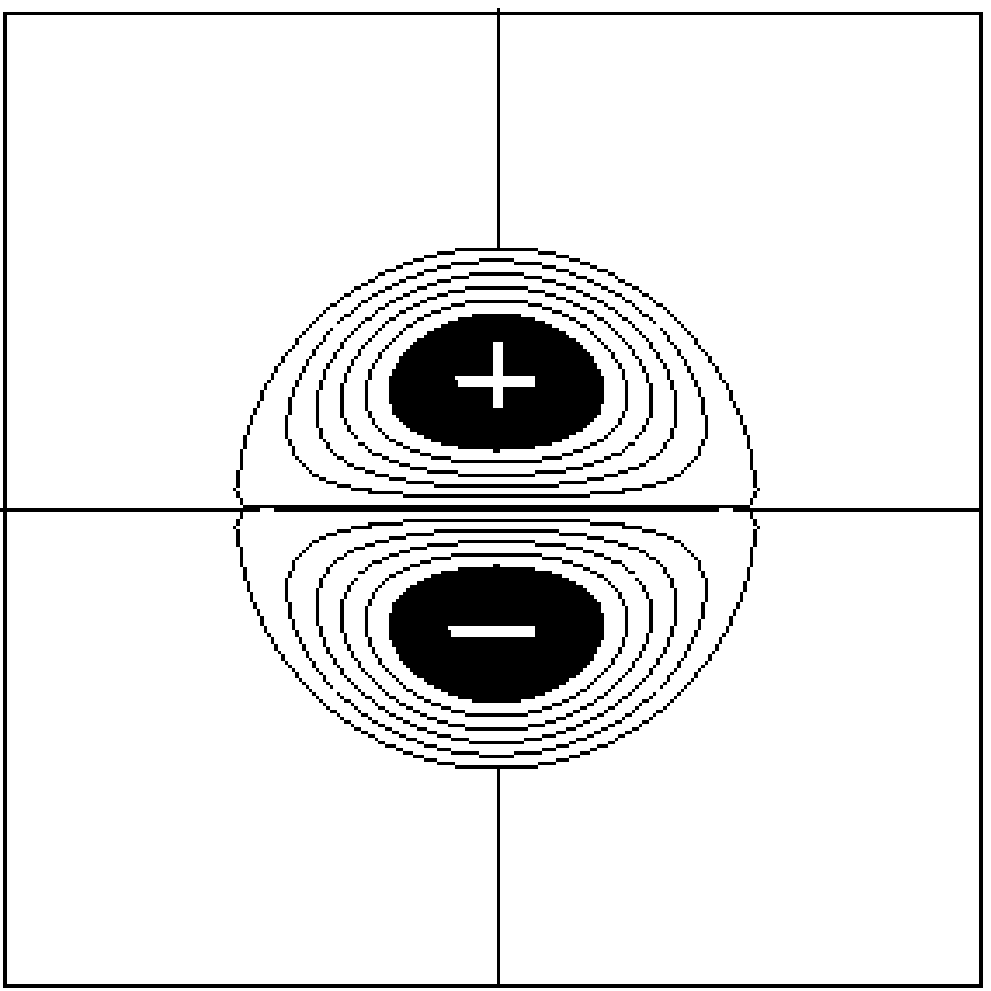}}
\end{minipage}
\hspace{0.5in}
\begin{minipage}[c]{.476 \linewidth}
\scalebox{1}[1.12]{\includegraphics[width=\linewidth]{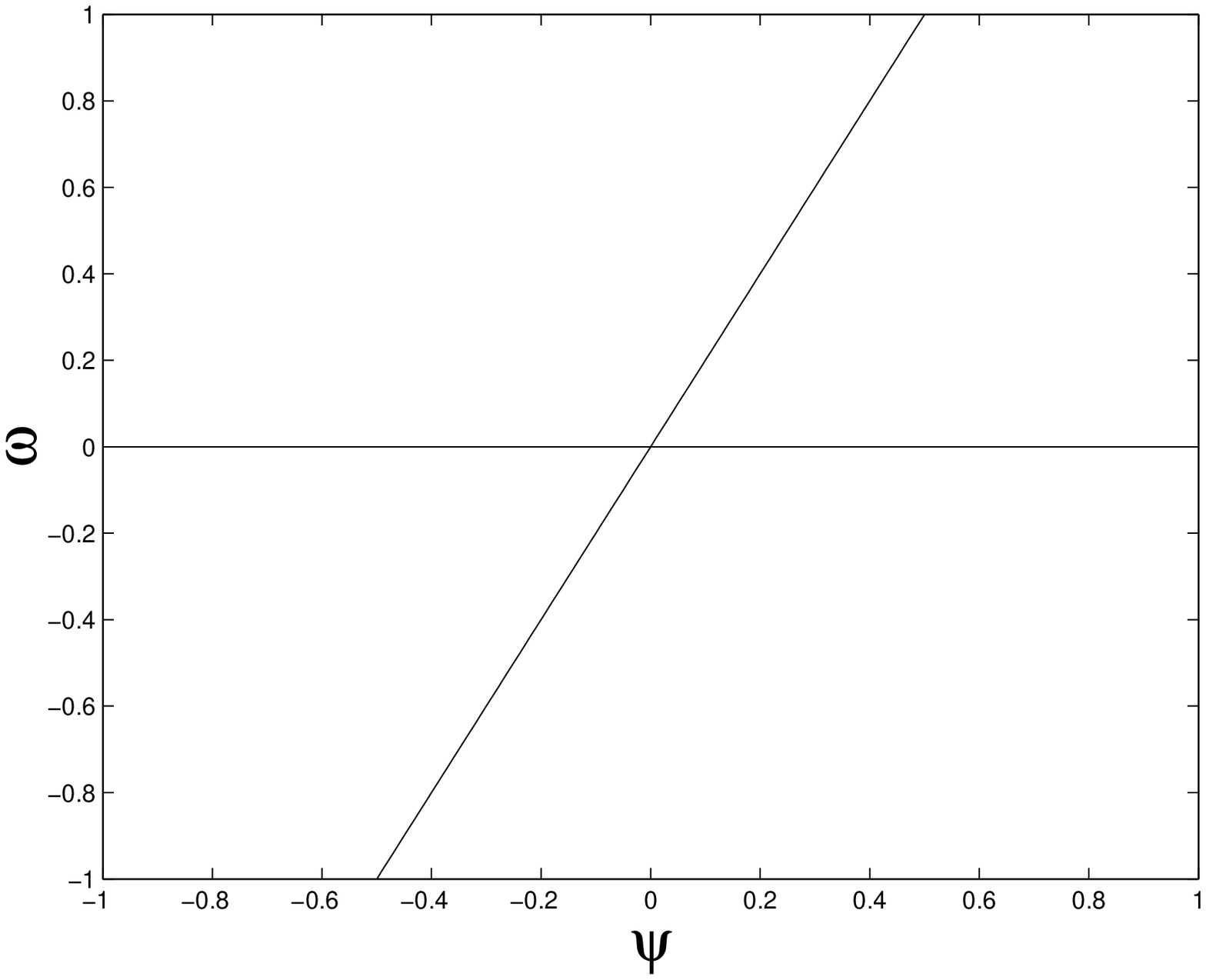}}
\end{minipage}
\caption{ The Lamb dipole with a linear relationship between
$\omega$ and $\psi$, can be considered as a stationary
two-dimensional solution of the Euler equation. The figure on the
left indicates the vorticity field, with one positive vortex and
one negative vortex confined within the circle. Outside that
circle, the vorticity is zero.} \label{fig:chaprea}
\end{figure*}

\begin{figure*}[!htbp]
\centering
\begin{minipage}[c]{.478 \linewidth}
\scalebox{1}[1]{\includegraphics[width=\linewidth]{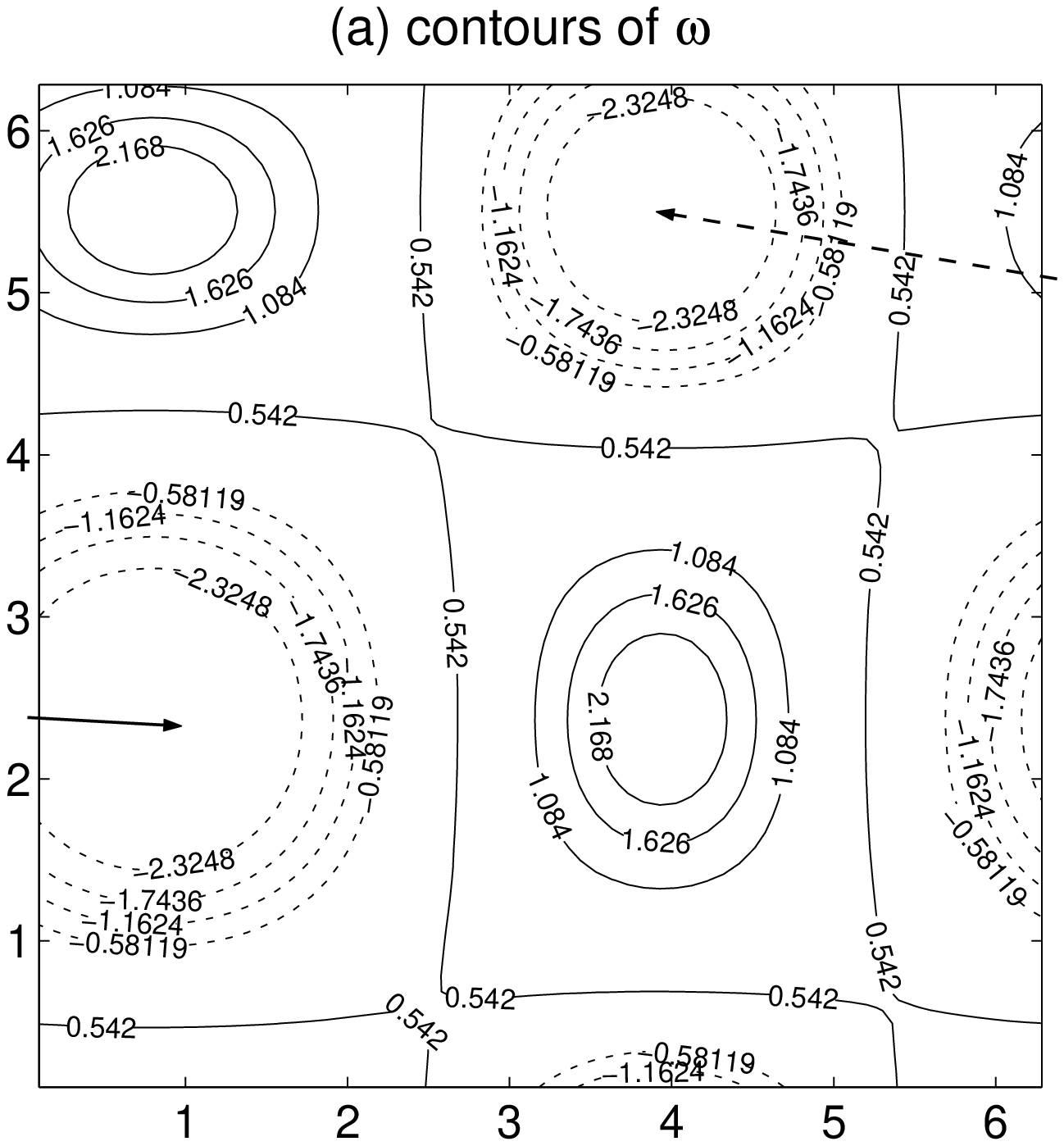}}
\end{minipage}
\begin{minipage}[c]{.478 \linewidth}
\scalebox{1}[1]{\includegraphics[width=\linewidth]{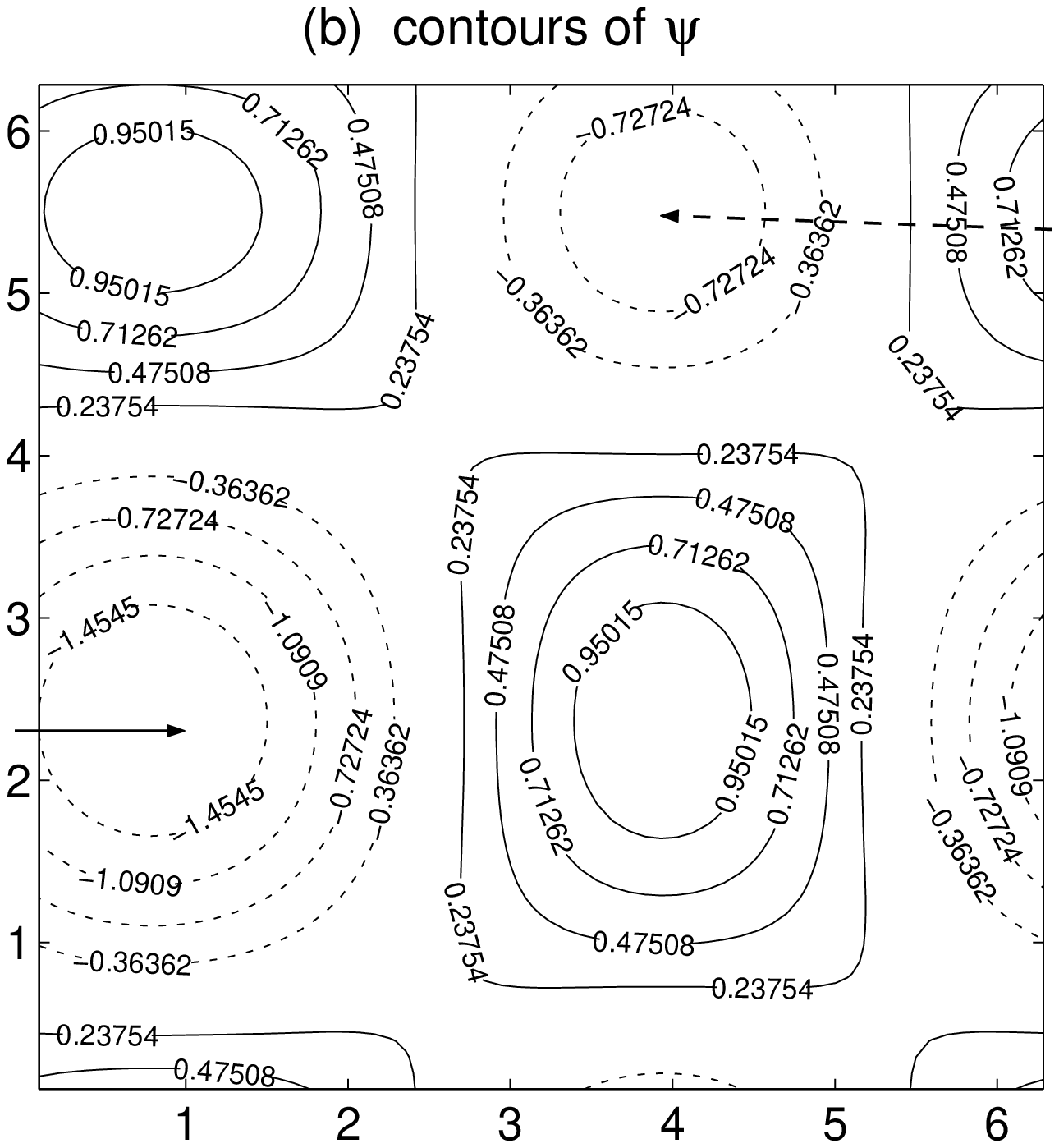}}
\end{minipage}
\begin{minipage}[c]{.68 \linewidth}
\scalebox{1}[1.1]{\includegraphics[width=\linewidth]{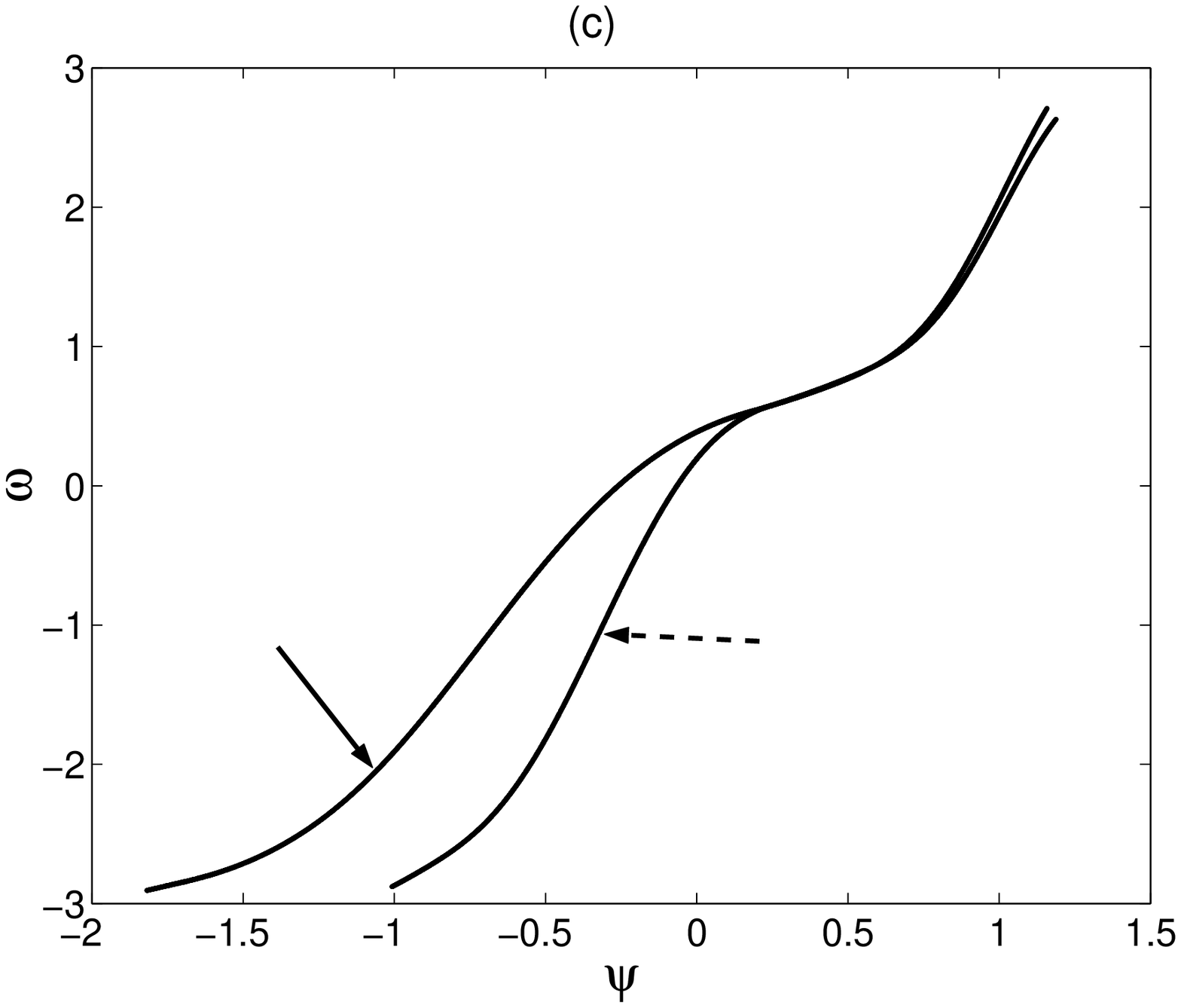}}
\end{minipage}
\caption{ The double-valued structure, representing a
non-functional $\omega-\psi$ relation (c), still is a stationary
solution of the Euler equation. The associated vorticity and
stream function contour plots are shown in (a) and (b).}
\label{fig:chapreb1}
\end{figure*}

\begin{figure*}[!htbp]
\centering \subfigure[]{\begin{minipage}[c]{.5 \linewidth}
\scalebox{1.2}[1]{\includegraphics[width=\linewidth]{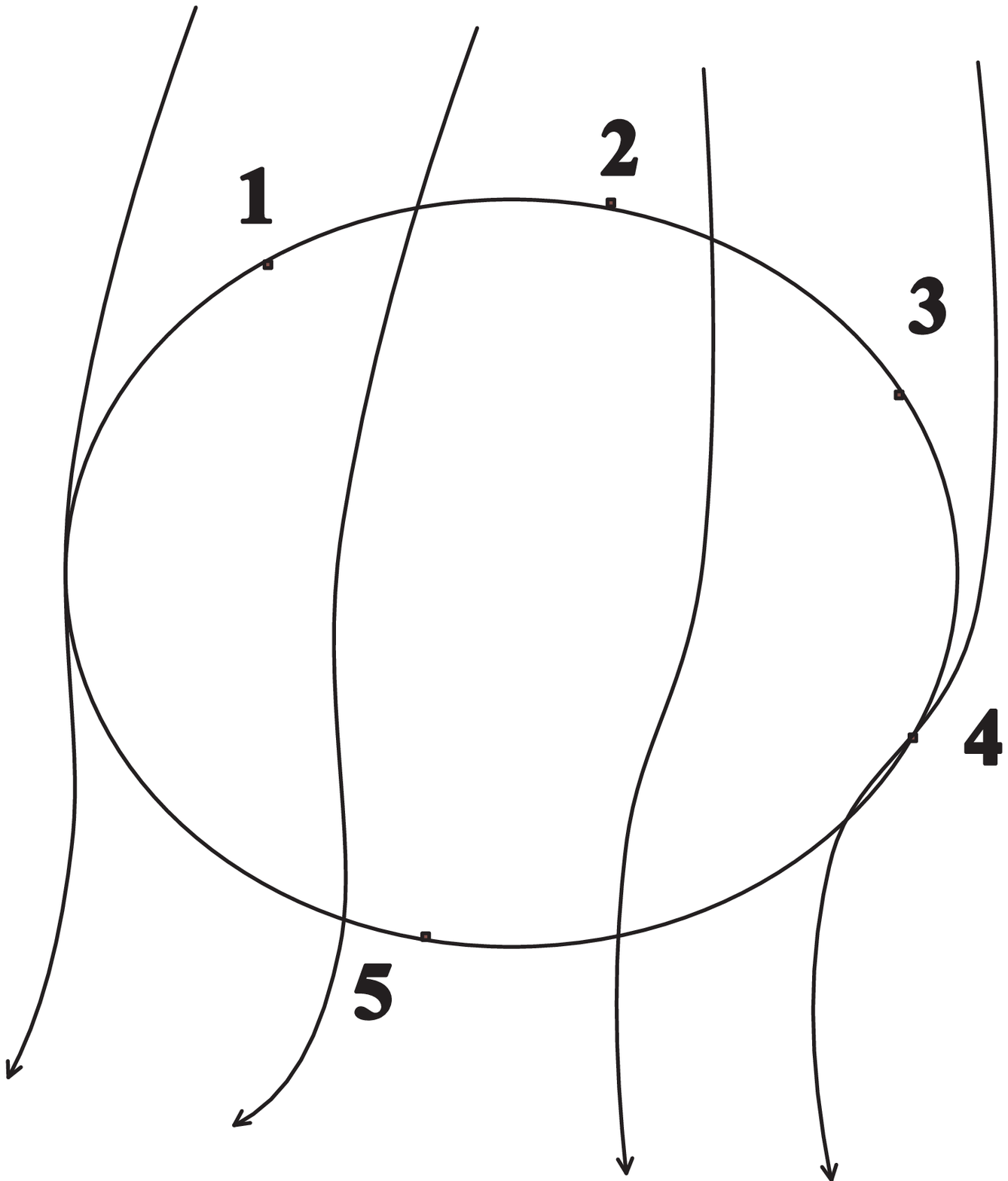}}
\end{minipage}}

\subfigure[]{\begin{minipage}[c]{.45 \linewidth}
\scalebox{1}[1]{\includegraphics[width=\linewidth]{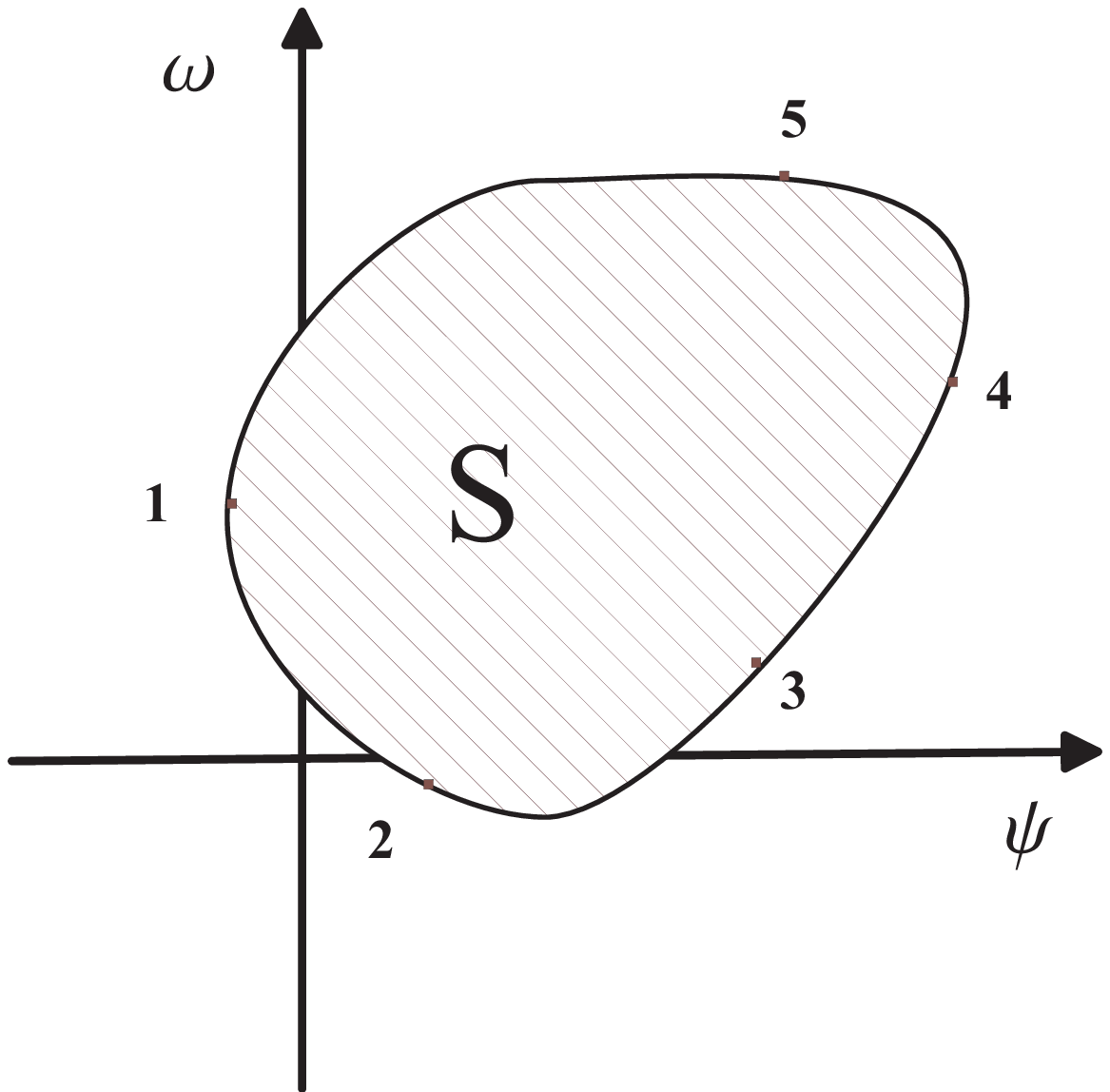}}
\end{minipage}}
\subfigure[]{\begin{minipage}[c]{.45 \linewidth}
\scalebox{1}[1]{\includegraphics[width=\linewidth]{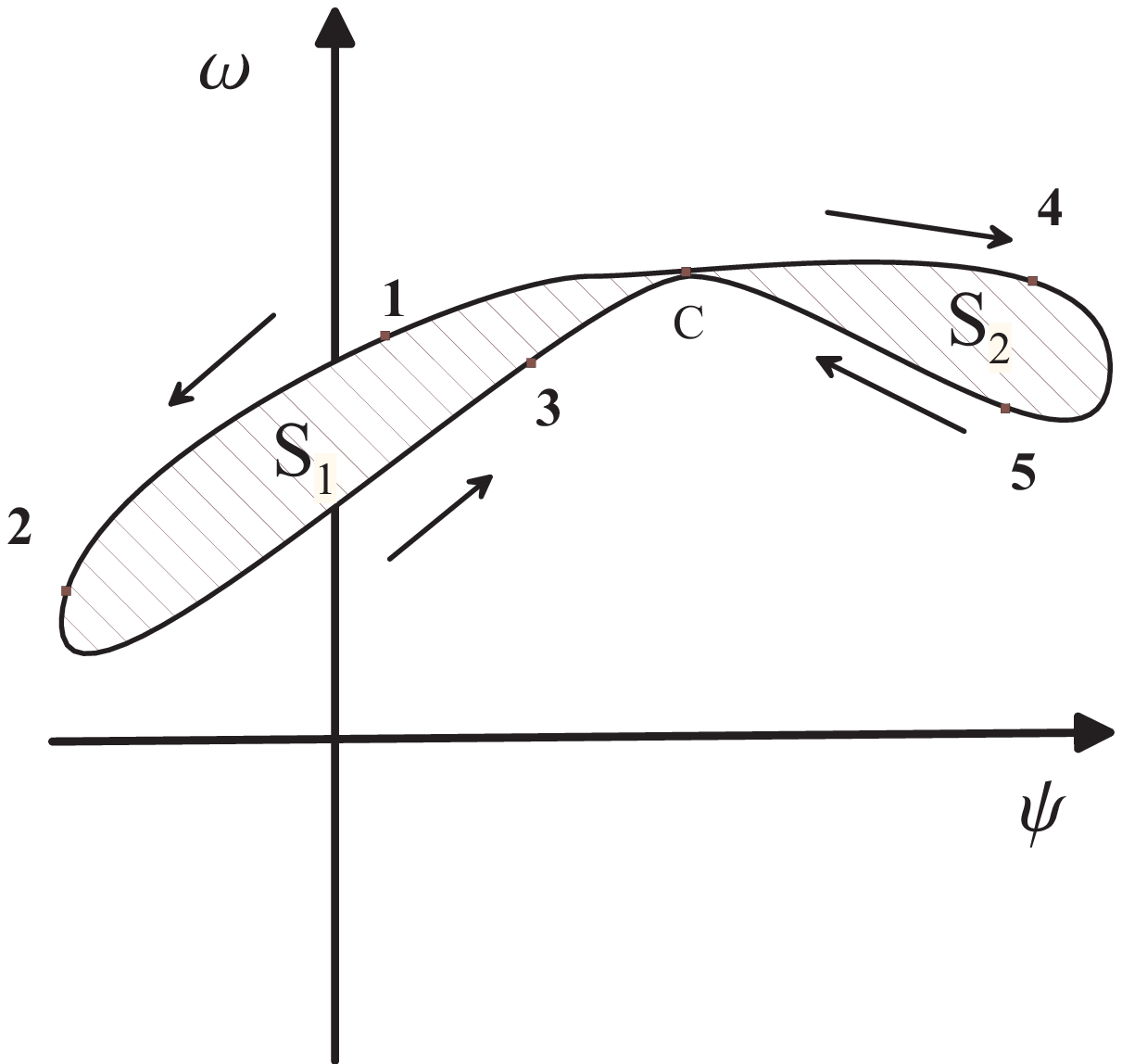}}
\end{minipage}}
\caption{ (a) is the contour plot of the stream function $\psi$
for a flow field, there is a circuit with five marked points on
it; (b) indicates one possible distribution of the five marked
points in the $\omega-\psi$ space - they form a simple circuit;
(c) indicates another possibility - a reentrant area: $S_1$ is the
anticlockwise region and $S_2$ the clockwise region.}
\label{fig:suite3}
\end{figure*}

\begin{figure*}[!htbp]
\centering \subfigure[]{\begin{minipage}[c]{.45 \linewidth}
\scalebox{1}[1]{\includegraphics[width=\linewidth]{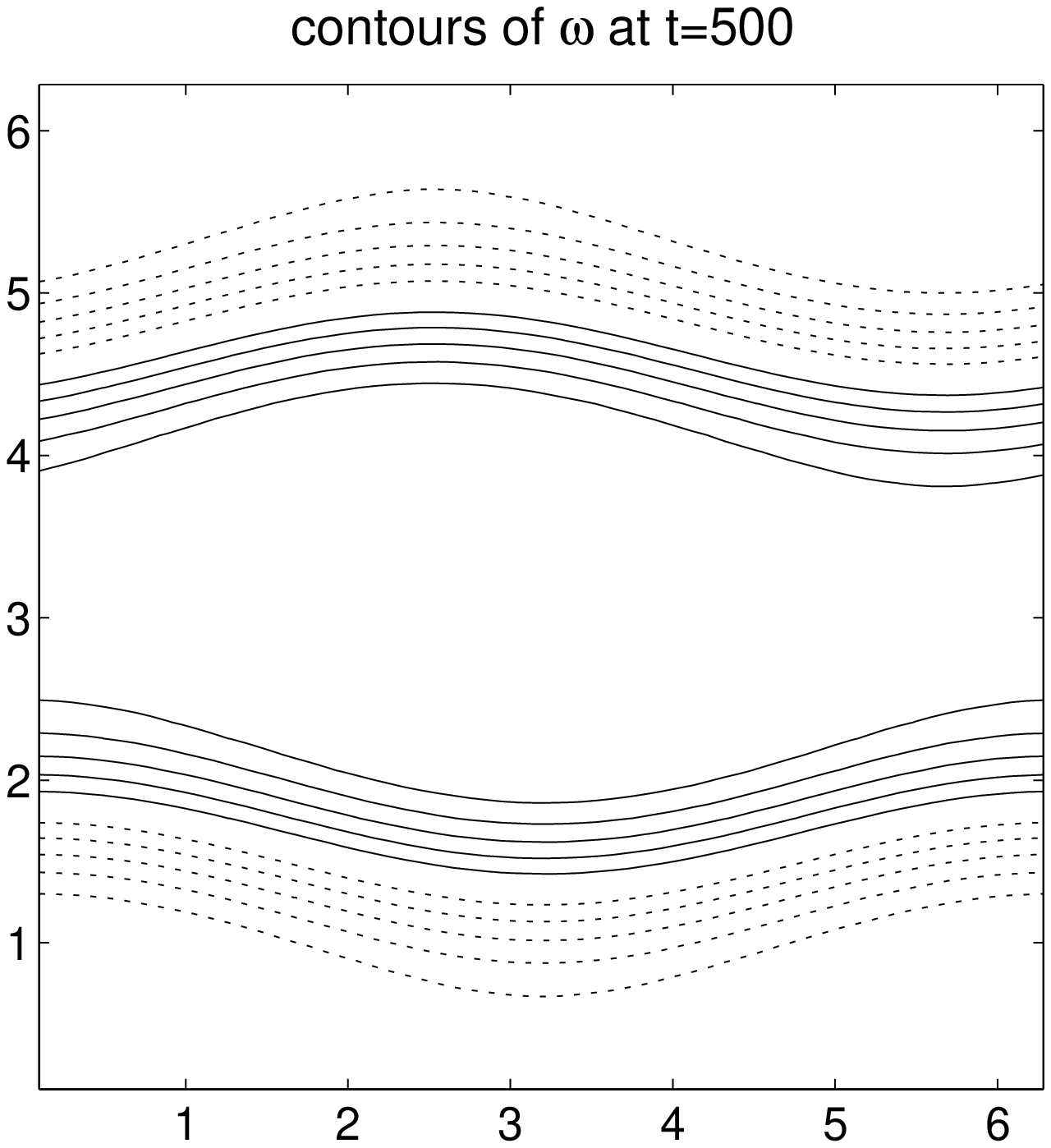}}
\end{minipage}}
\subfigure[]{\begin{minipage}[c]{.52 \linewidth}
\scalebox{1}[1.1]{\includegraphics[width=\linewidth]{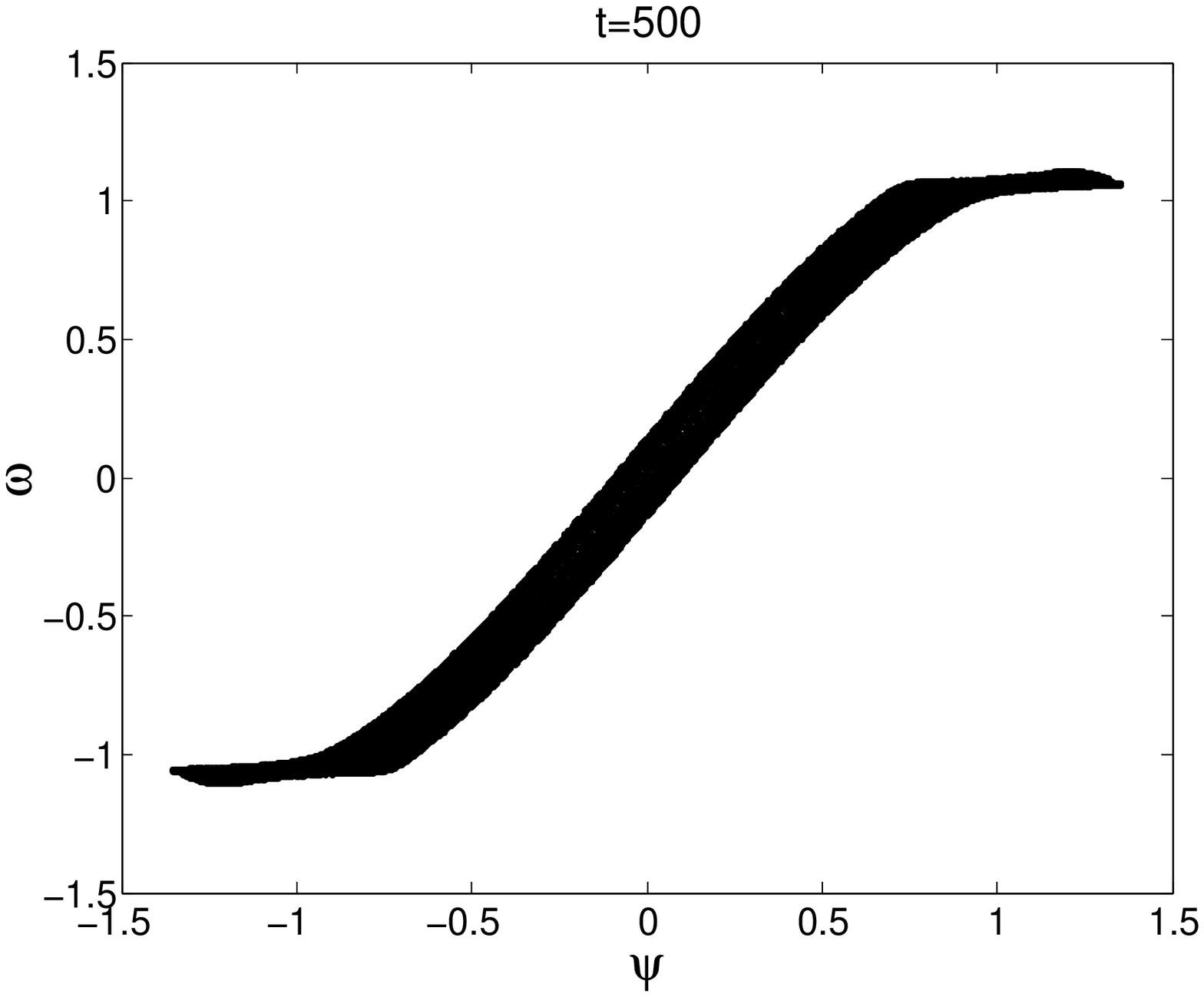}}
\end{minipage}}
\caption{In the late stage of the quadrupole to bar simulation
[3], a traveling wave appears and exists for very long time: from
$t=50$ to $t=1000$. (a) shows one contour plot of the vorticity
during this stage; (b) shows the corresponding $\omega-\psi$ plot.
The points in (b) cover a band of area that cannot be treated as
zero.} \label{fig:newps1}
\end{figure*}

\begin{figure*}[!htbp]
\centering
\begin{minipage}[c]{.3 \linewidth}
\scalebox{1}[.9]{\includegraphics[width=\linewidth]{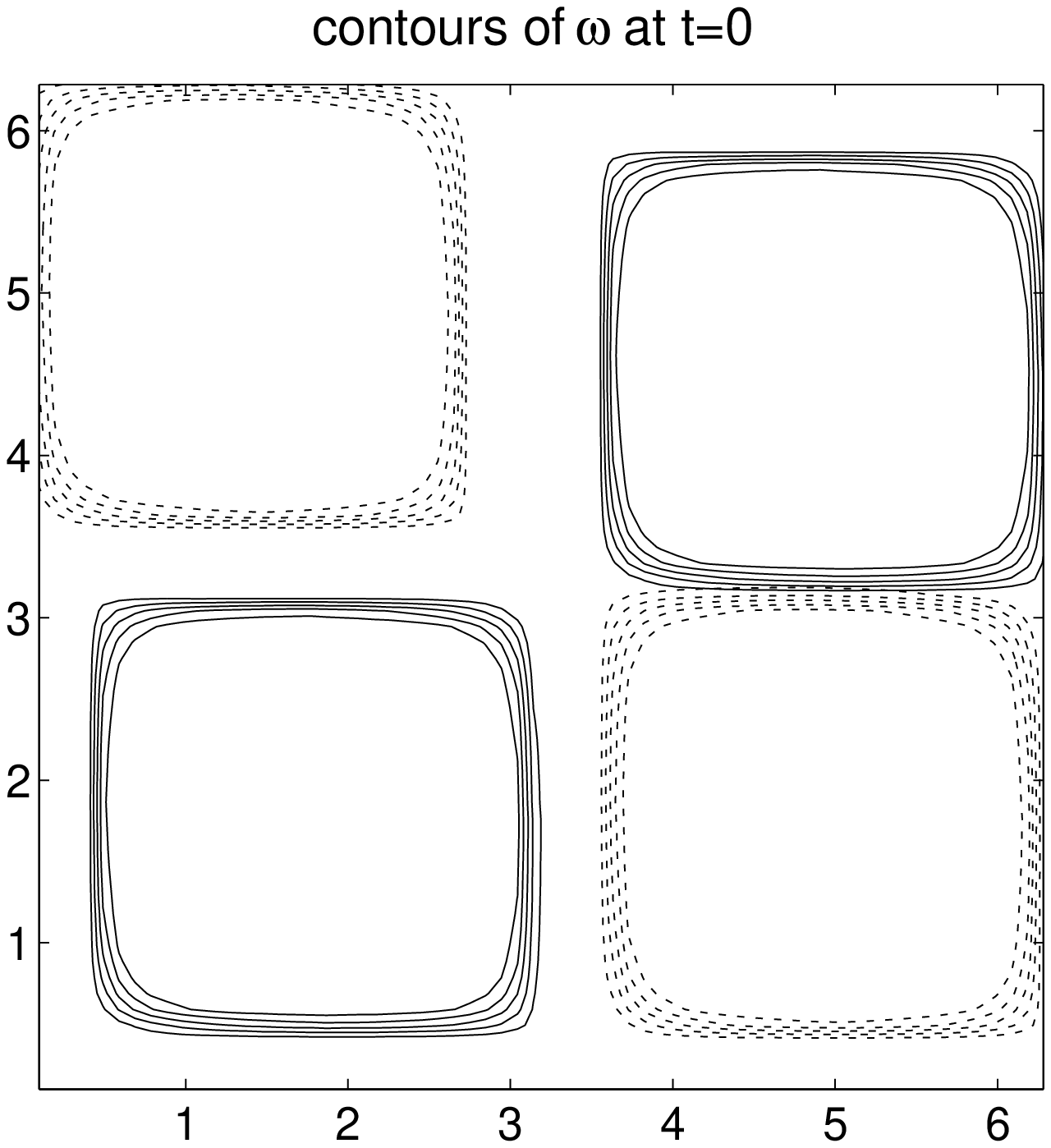}}
\end{minipage}
\hspace{0.65in}
\begin{minipage}[c]{.3 \linewidth}
\scalebox{1}[.9]{\includegraphics[width=\linewidth]{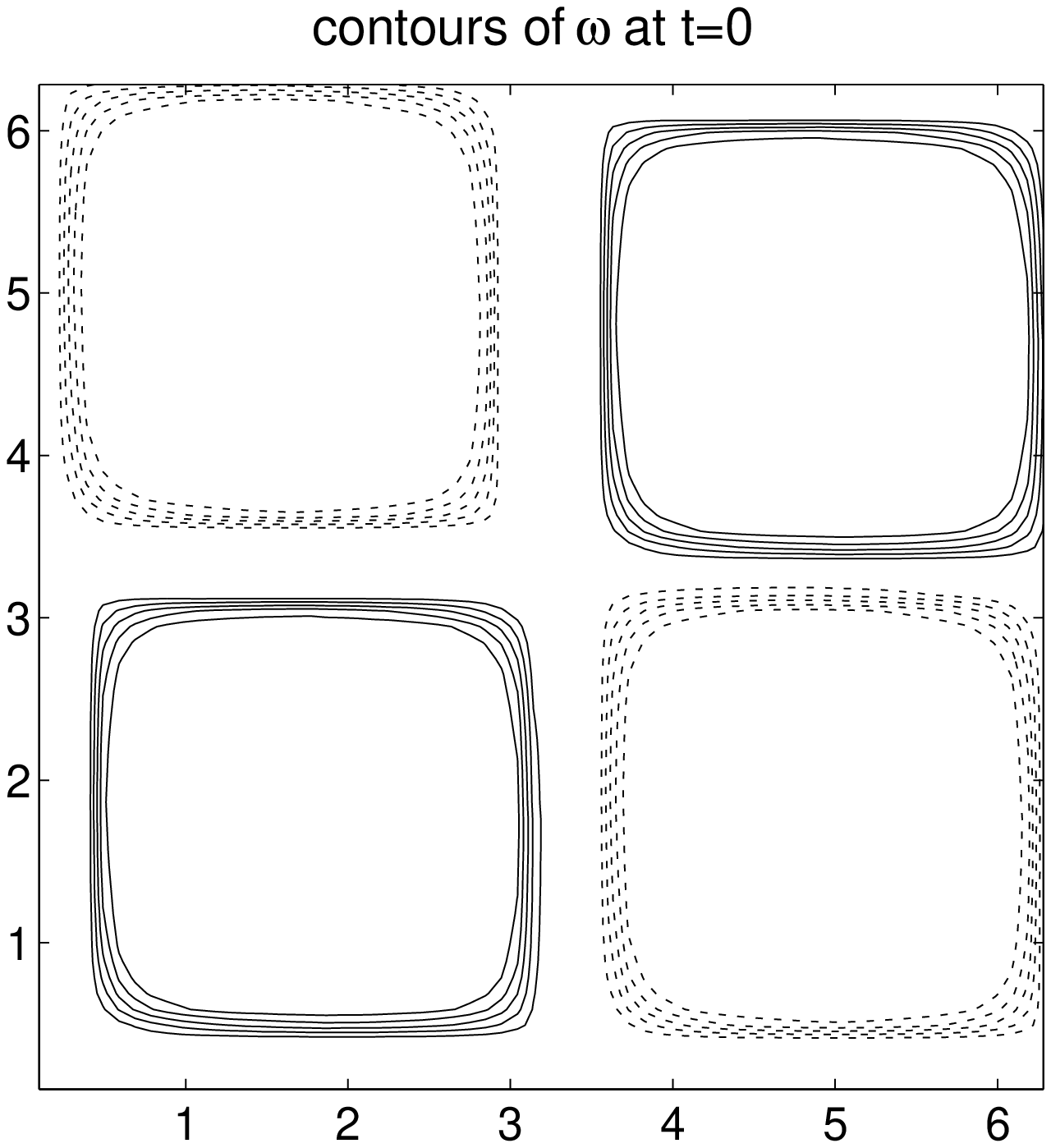}}
\end{minipage}
\begin{minipage}[c]{.3 \linewidth}
\scalebox{1}[.9]{\includegraphics[width=\linewidth]{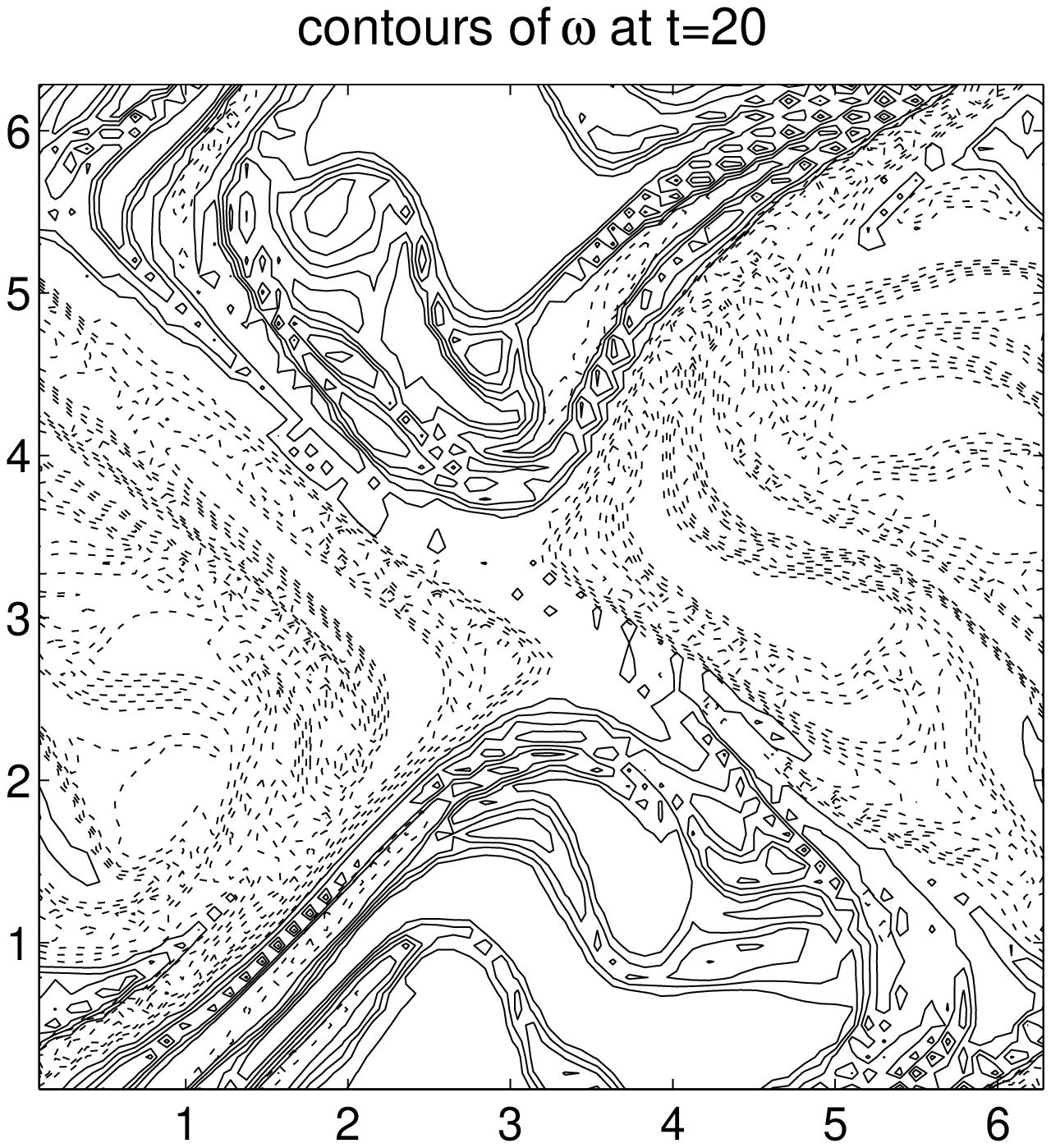}}
\end{minipage}
\hspace{0.65in}
\begin{minipage}[c]{.3 \linewidth}
\scalebox{1}[.9]{\includegraphics[width=\linewidth]{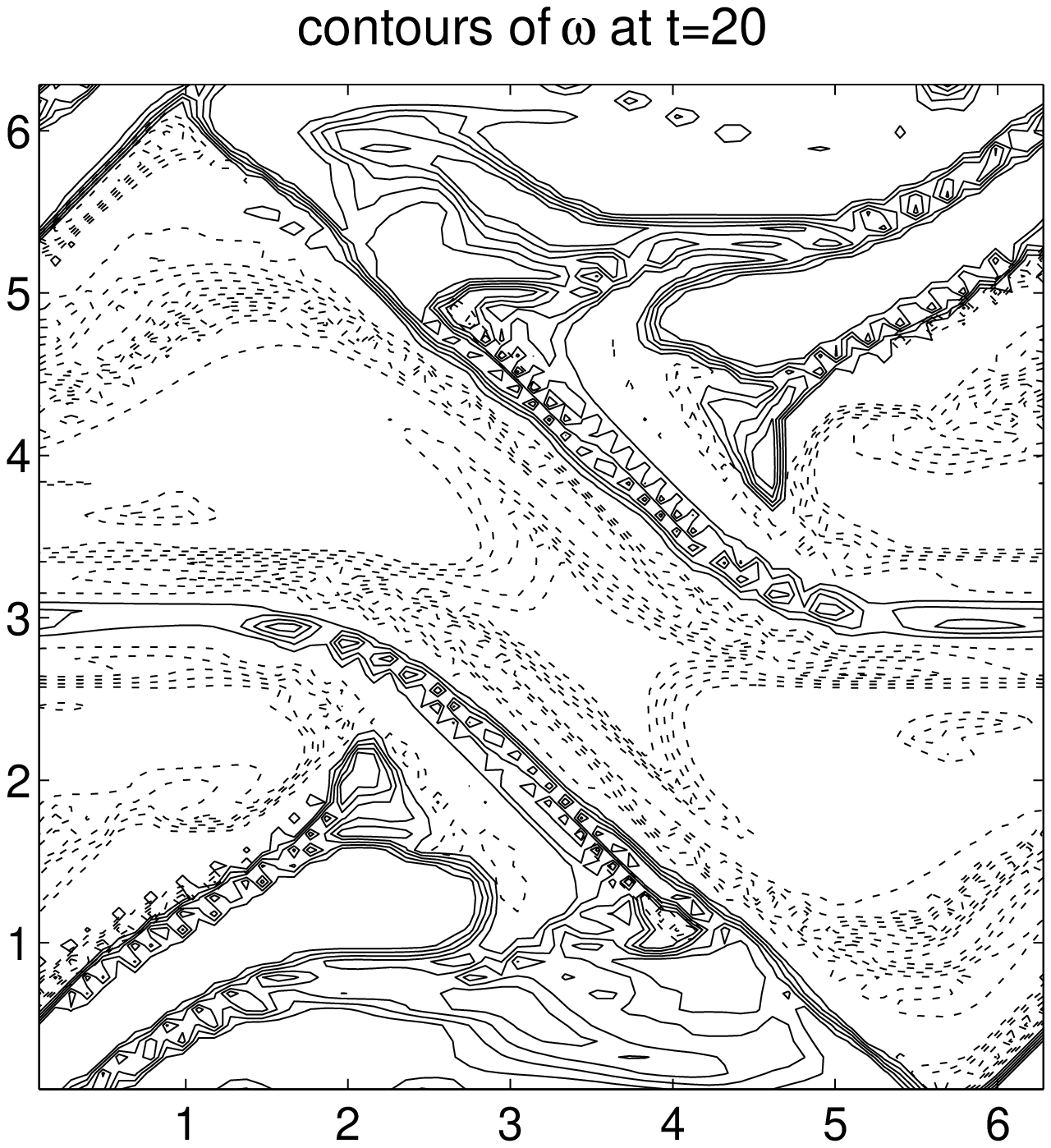}}
\end{minipage}
\begin{minipage}[c]{.3 \linewidth}
\scalebox{1}[.9]{\includegraphics[width=\linewidth]{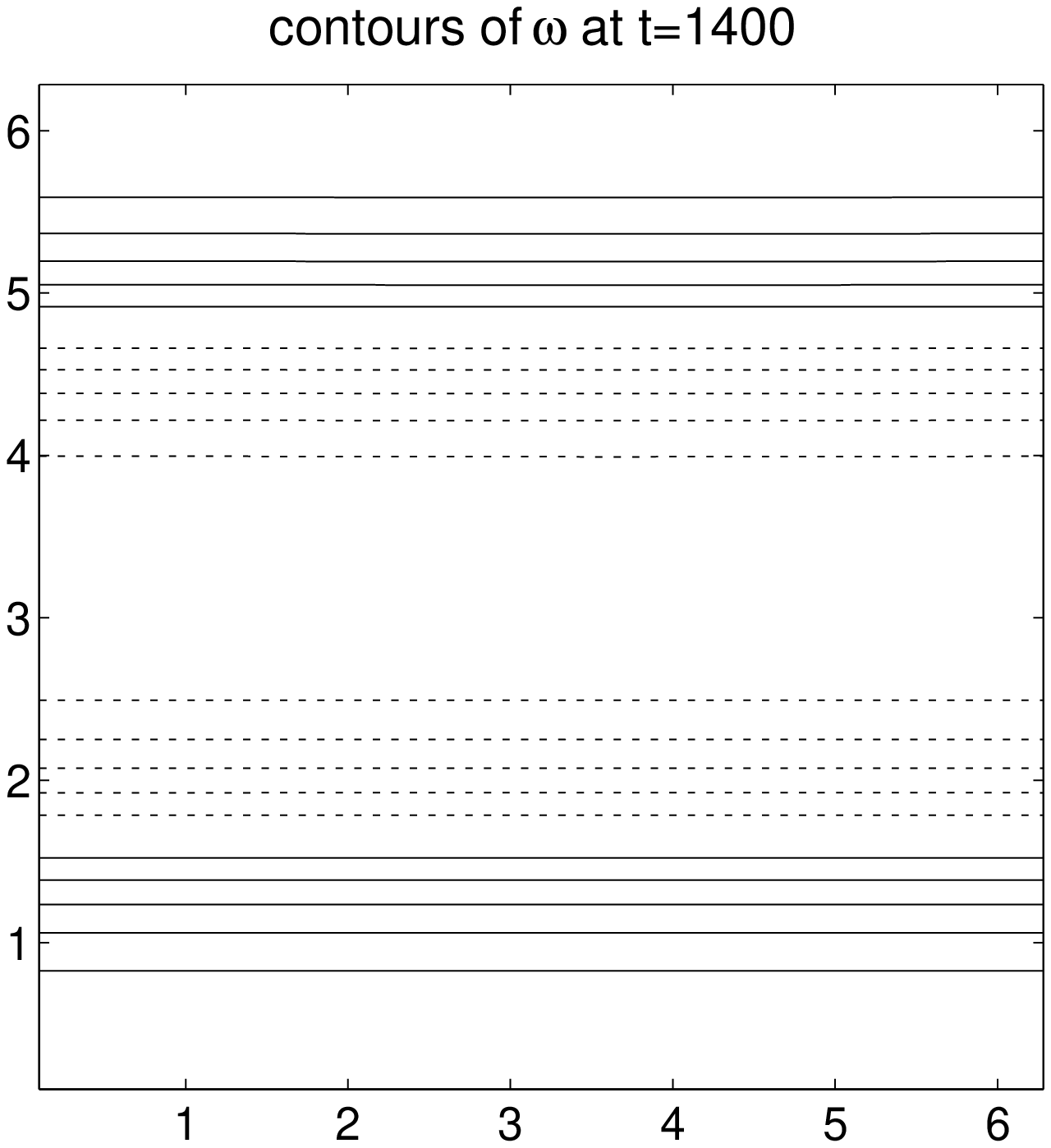}}
\end{minipage}
\hspace{0.65in}
\begin{minipage}[c]{.3 \linewidth}
\scalebox{1}[.9]{\includegraphics[width=\linewidth]{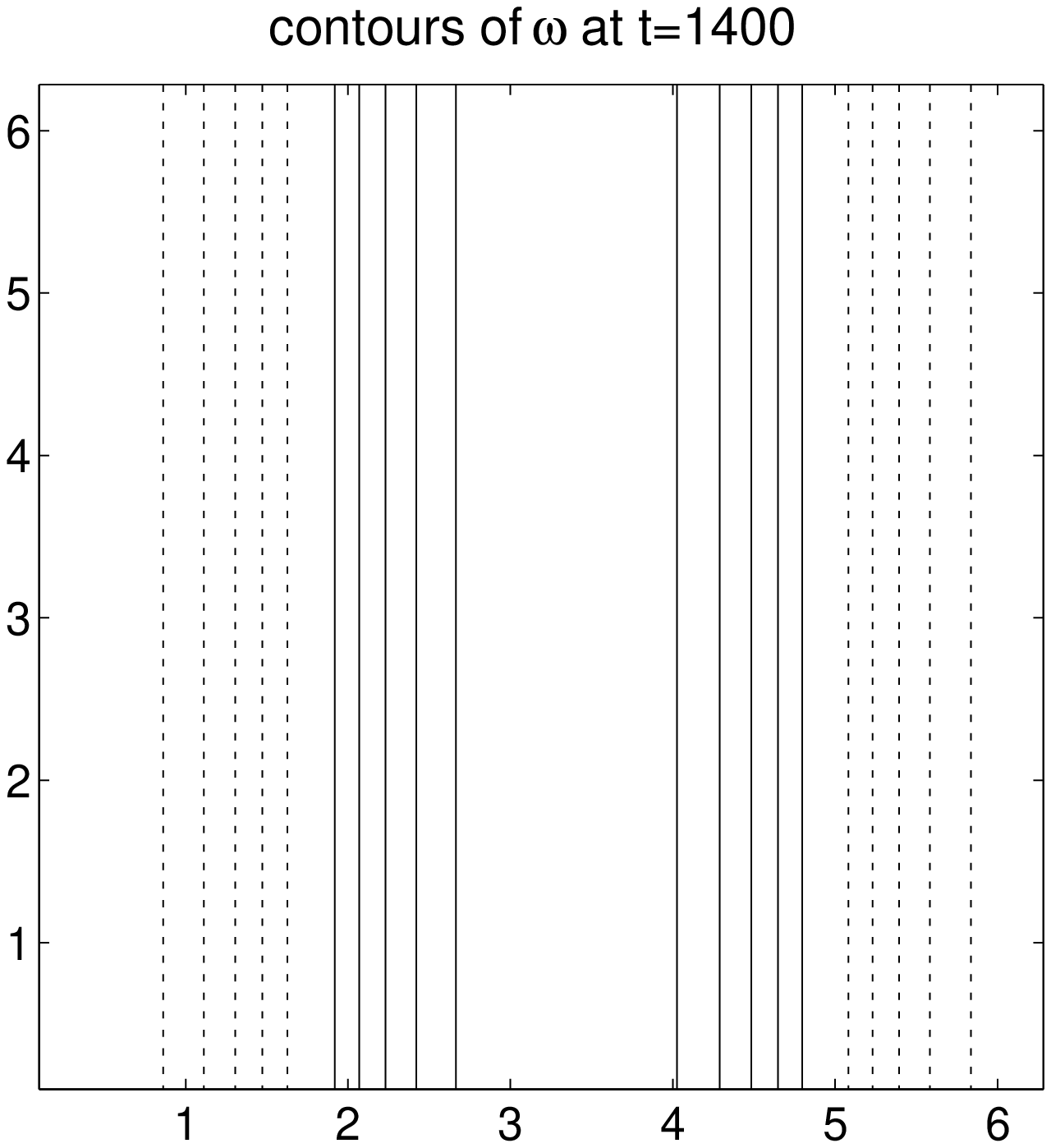}}
\end{minipage}
\subfigure[]{
\begin{minipage}[c]{.3 \linewidth}
\scalebox{1}[.9]{\includegraphics[width=\linewidth]{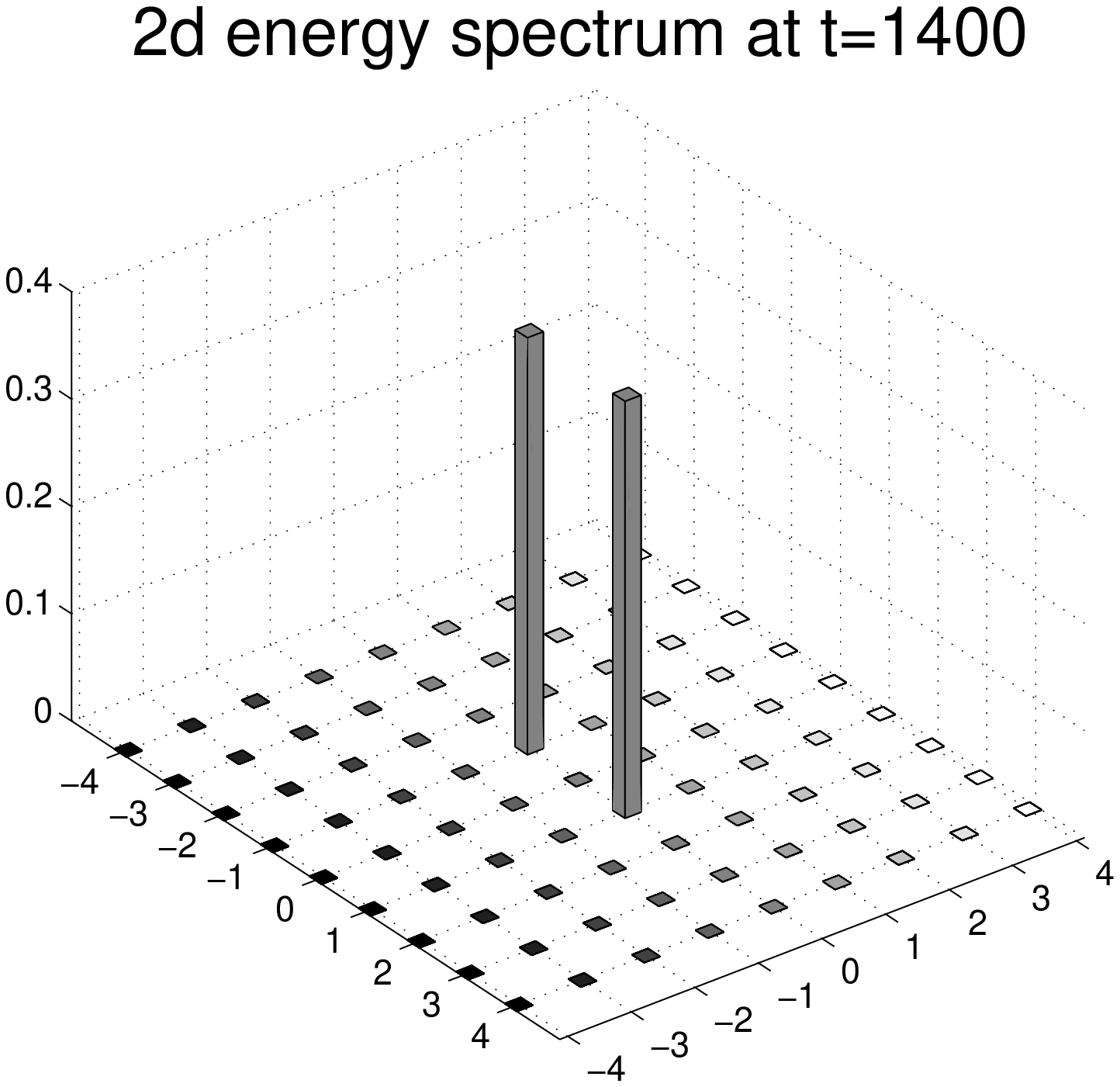}}
\end{minipage}}
\hspace{0.65in} \subfigure[]{
\begin{minipage}[c]{.3 \linewidth}
\scalebox{1}[.9]{\includegraphics[width=\linewidth]{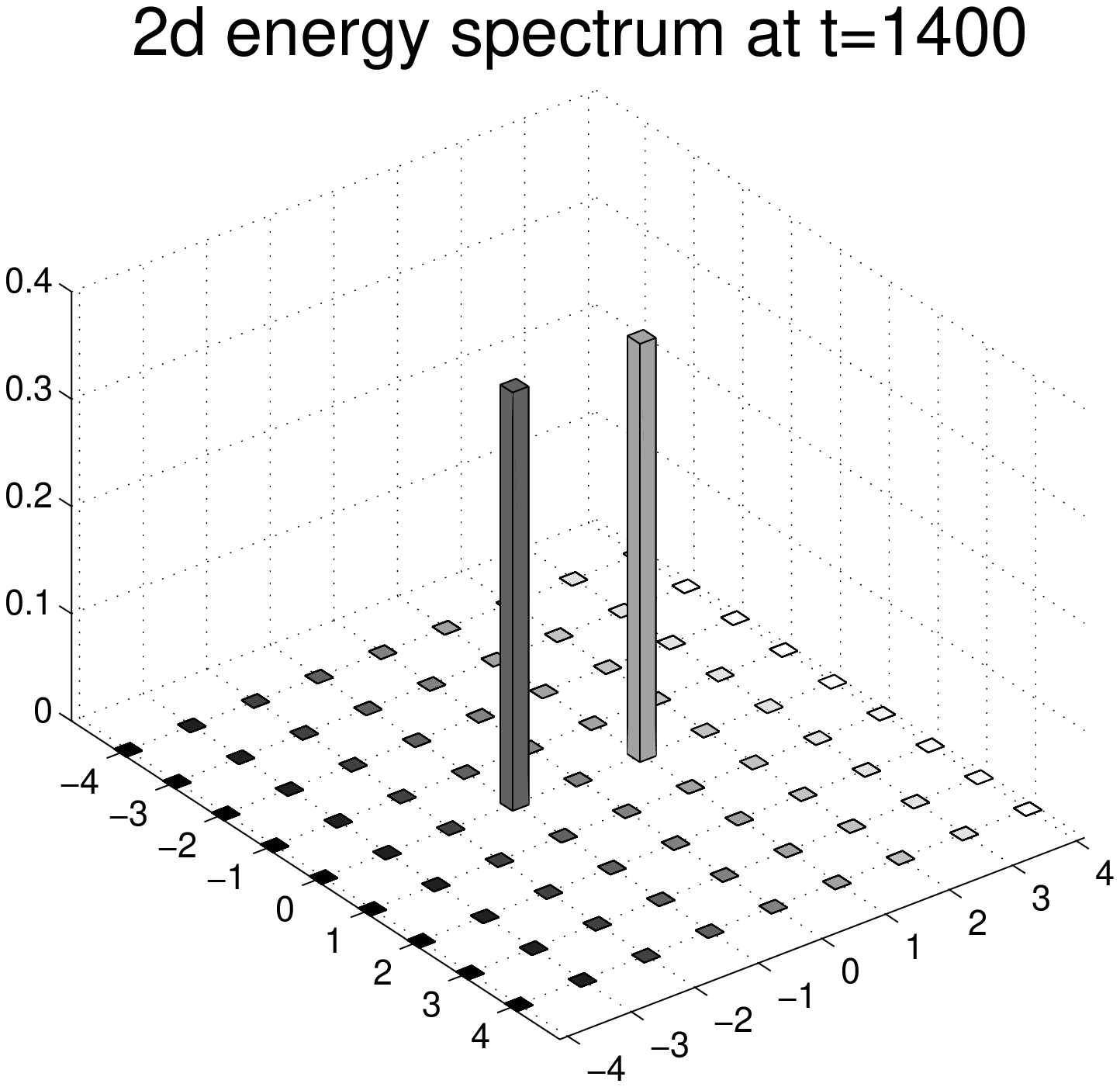}}
\end{minipage}}
\caption{The first three rows are contours of constant vorticity
for two runs with slightly different initial conditions in the
left (a) and the right (b) column. In both runs, the initial patch
sizes are reduced by a factor of $7/8 \times 7/8 = 49/64$ , and
the patches are displaced with respect to the quadrupole initial
condition shown in Fig. 7(a) of YMC. Pictures in the fourth row
are modal energies of final states at low wavenumbers for two
runs.} \label{fig:morebar1}
\end{figure*}

\begin{figure*}[!htbp]
\centering
\begin{minipage}[c]{.3 \linewidth}
\scalebox{1}[.9]{\includegraphics[width=\linewidth]{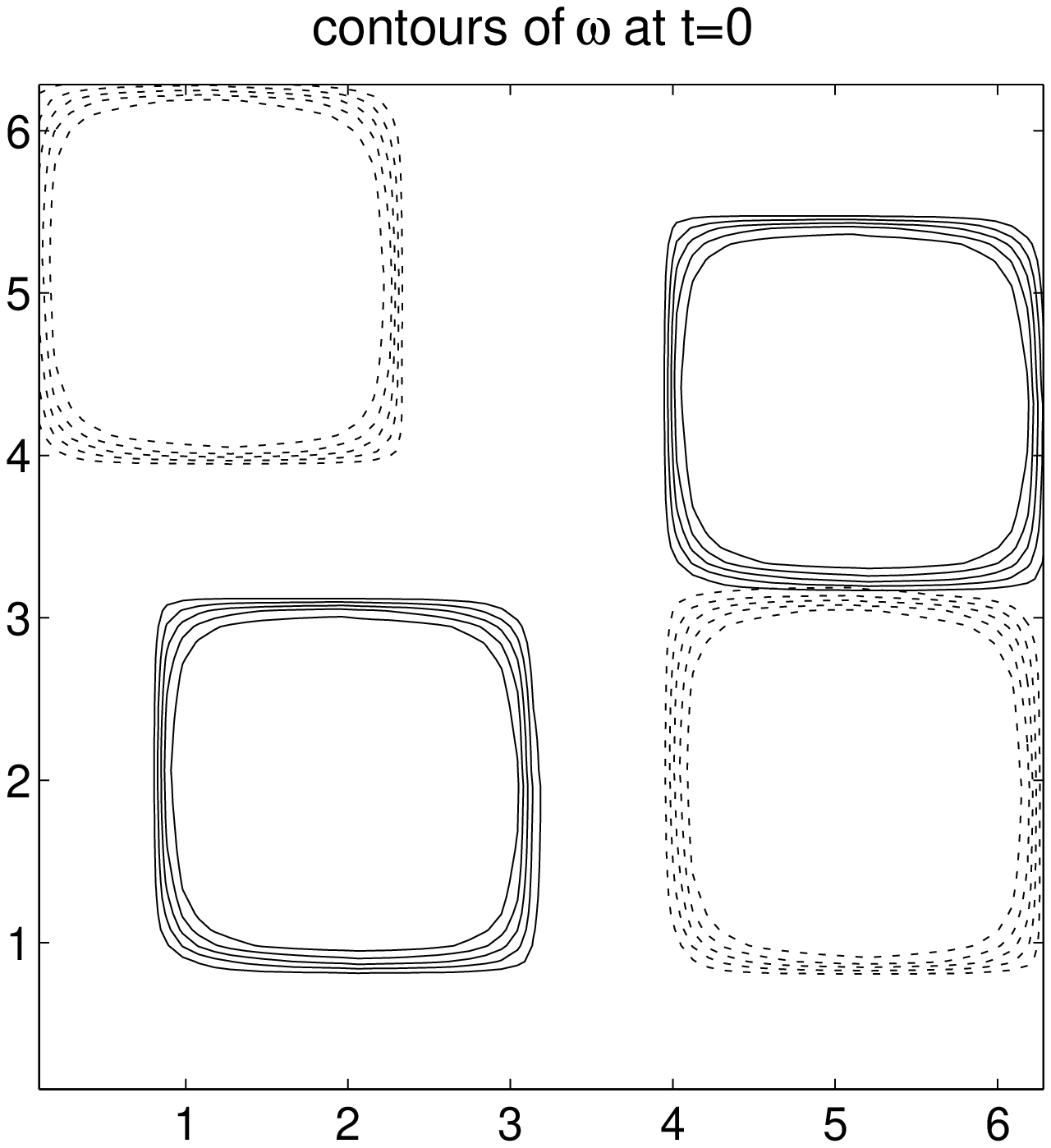}}
\end{minipage}
\hspace{0.65in}
\begin{minipage}[c]{.3 \linewidth}
\scalebox{1}[.9]{\includegraphics[width=\linewidth]{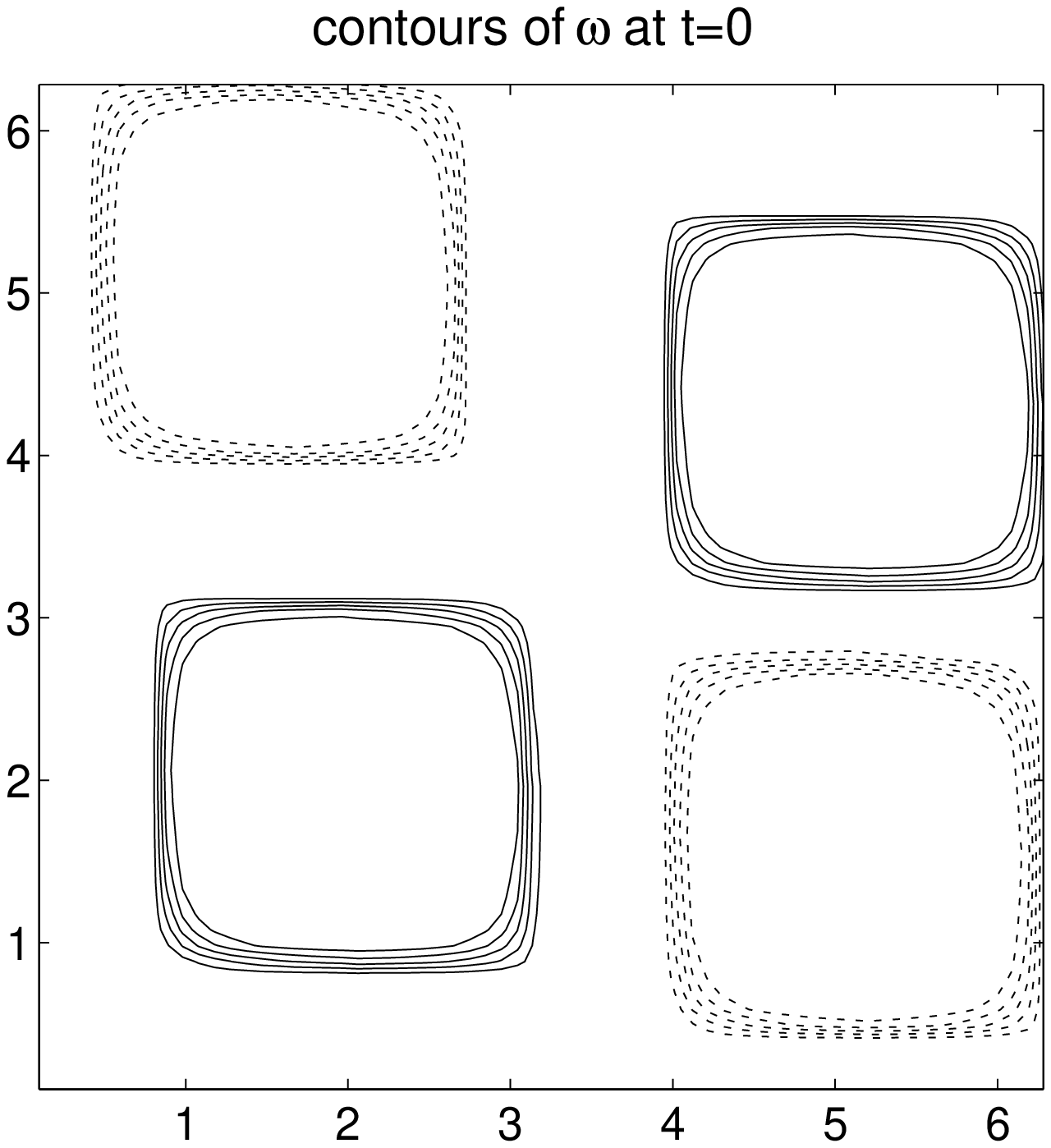}}
\end{minipage}
\begin{minipage}[c]{.3 \linewidth}
\scalebox{1}[.9]{\includegraphics[width=\linewidth]{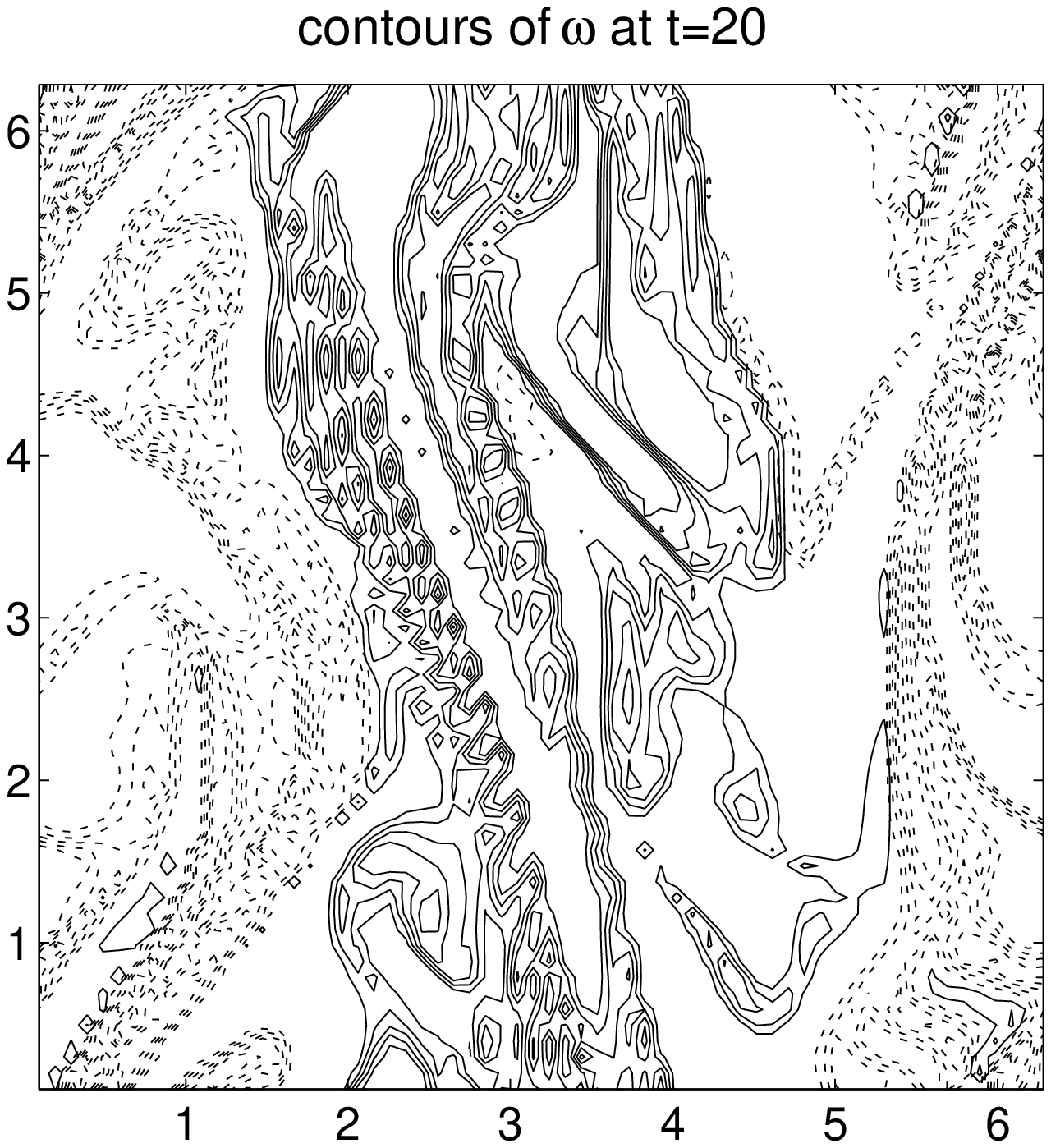}}
\end{minipage}
\hspace{0.65in}
\begin{minipage}[c]{.3 \linewidth}
\scalebox{1}[.9]{\includegraphics[width=\linewidth]{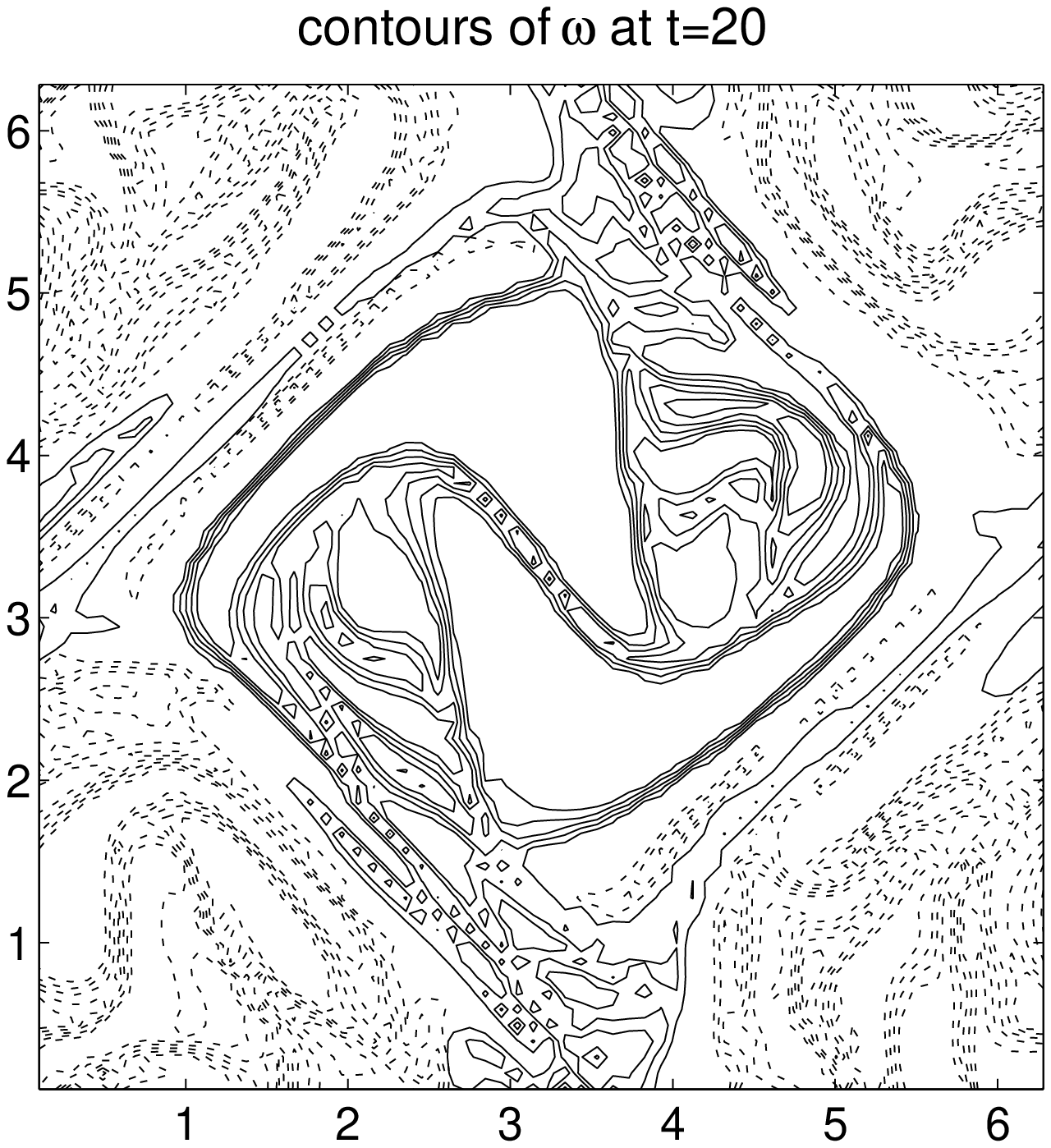}}
\end{minipage}
\begin{minipage}[c]{.3 \linewidth}
\scalebox{1}[.9]{\includegraphics[width=\linewidth]{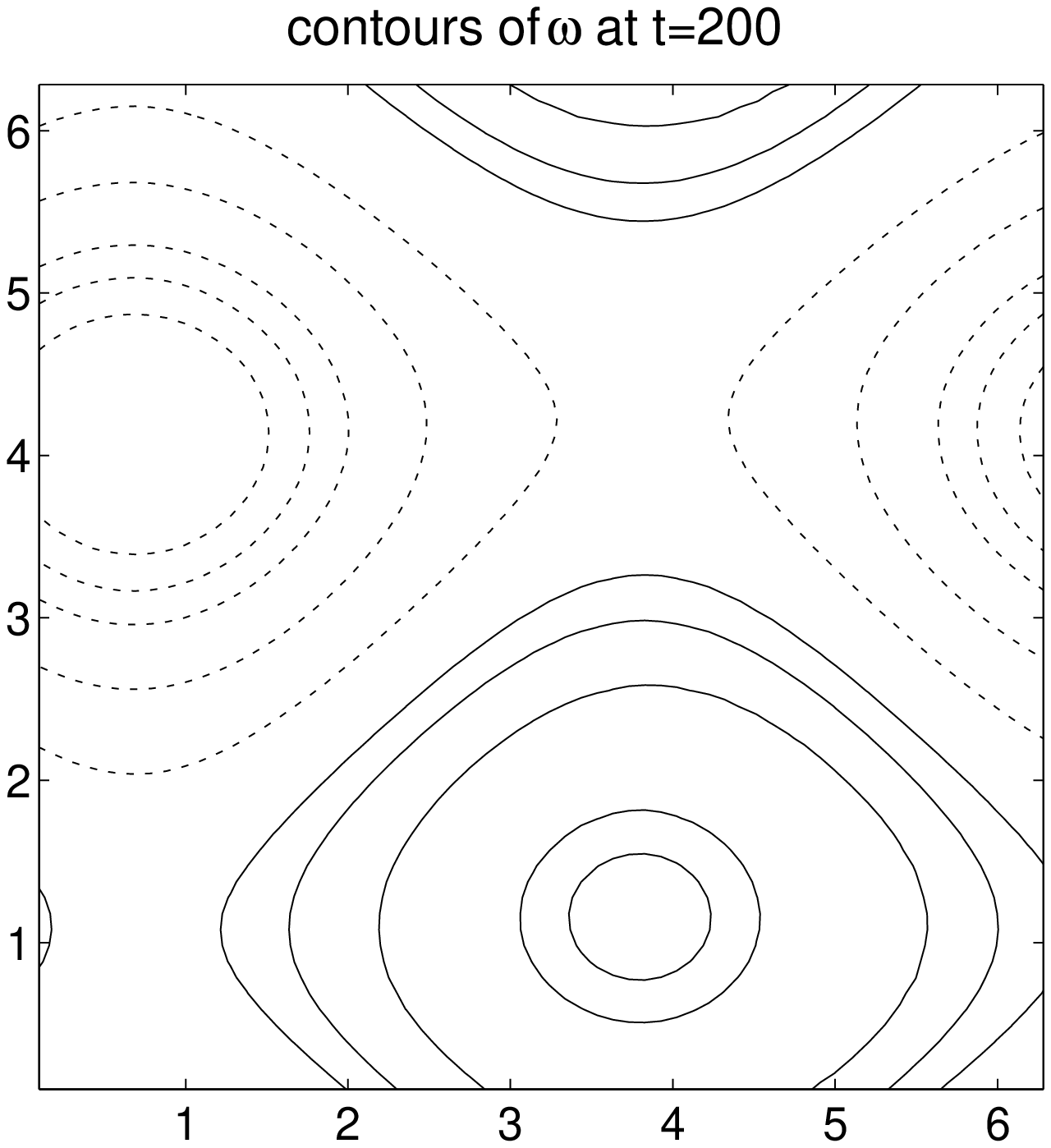}}
\end{minipage}
\hspace{0.65in}
\begin{minipage}[c]{.3 \linewidth}
\scalebox{1}[.9]{\includegraphics[width=\linewidth]{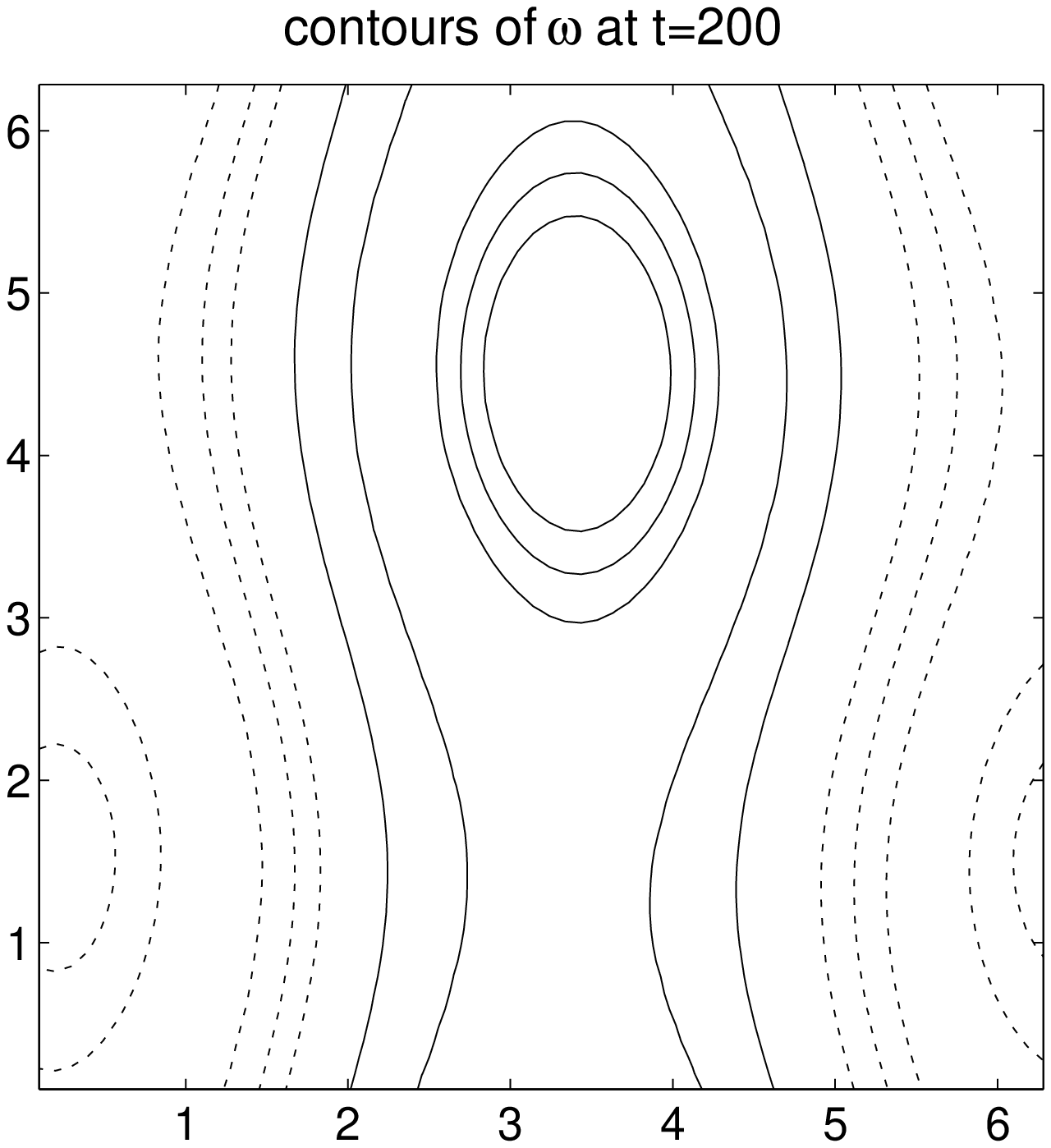}}
\end{minipage}
\subfigure[]{
\begin{minipage}[c]{.3 \linewidth}
\scalebox{1}[.9]{\includegraphics[width=\linewidth]{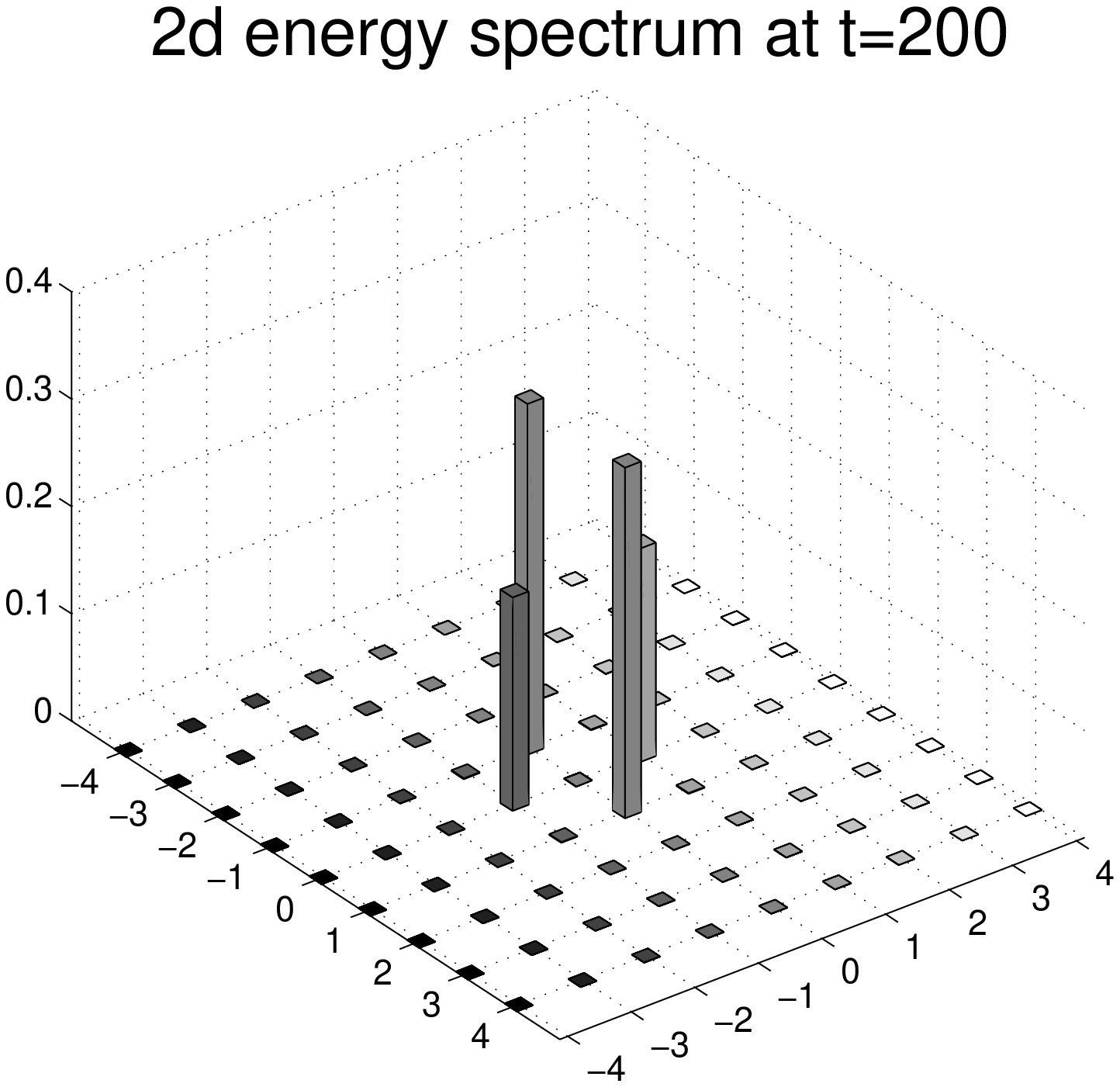}}
\end{minipage}}
\hspace{0.65in} \subfigure[]{
\begin{minipage}[c]{.3 \linewidth}
\scalebox{1}[.9]{\includegraphics[width=\linewidth]{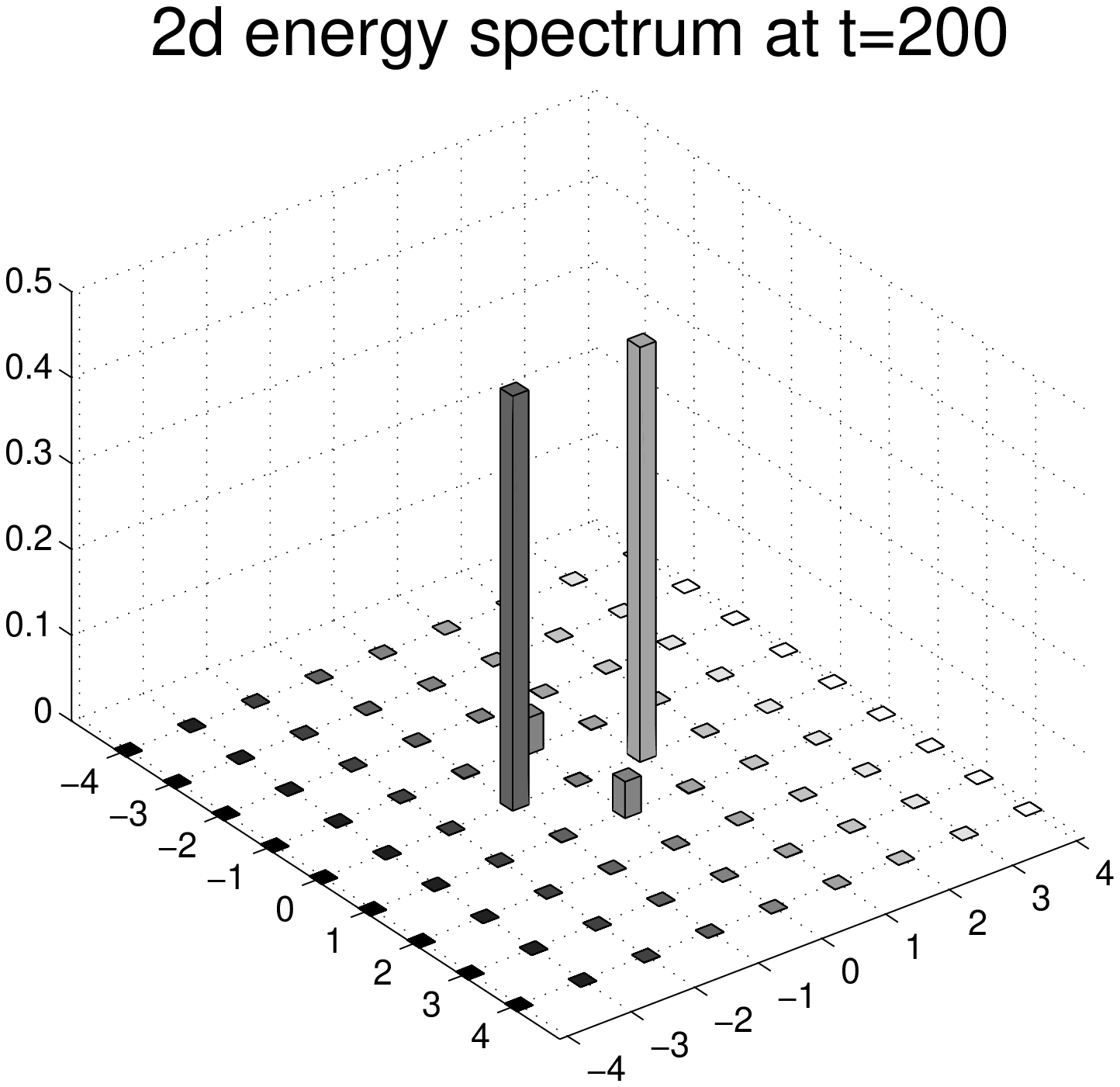}}
\end{minipage}}
\caption{ The first three rows are contours of constant vorticity
for two runs with slightly different initial conditions in the
left (a) and the right (b) column. In both runs, the initial patch
sizes are reduced by a factor of $3/4 \times 3/4 =9/16$, which are
somewhat larger reductions than initial conditions displayed in
Figs. 5, and the patches are displaced with respect to the
quadrupole initial condition shown in Fig. 7(a) of YMC. Pictures
in the fourth row are modal energies of final states at low
wavenumbers for two runs.} \label{fig:morebar2}
\end{figure*}

\begin{figure*}[!htbp]
\centering
\begin{minipage}[c]{.34 \linewidth}
\scalebox{1}[1.]{\includegraphics[width=\linewidth]{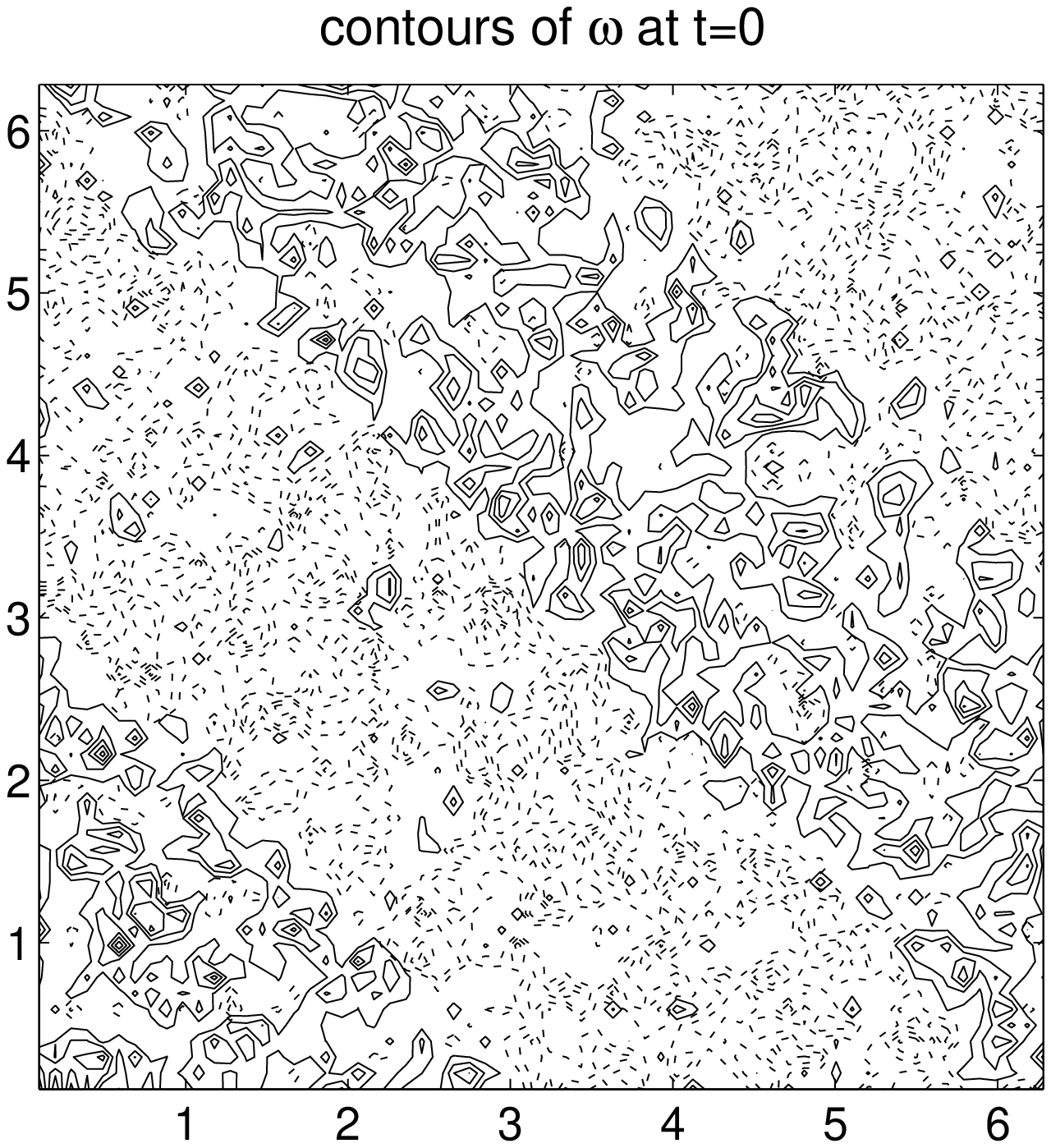}}
\end{minipage}
\hspace{0.5in}
\begin{minipage}[c]{.34 \linewidth}
\scalebox{1}[1.]{\includegraphics[width=\linewidth]{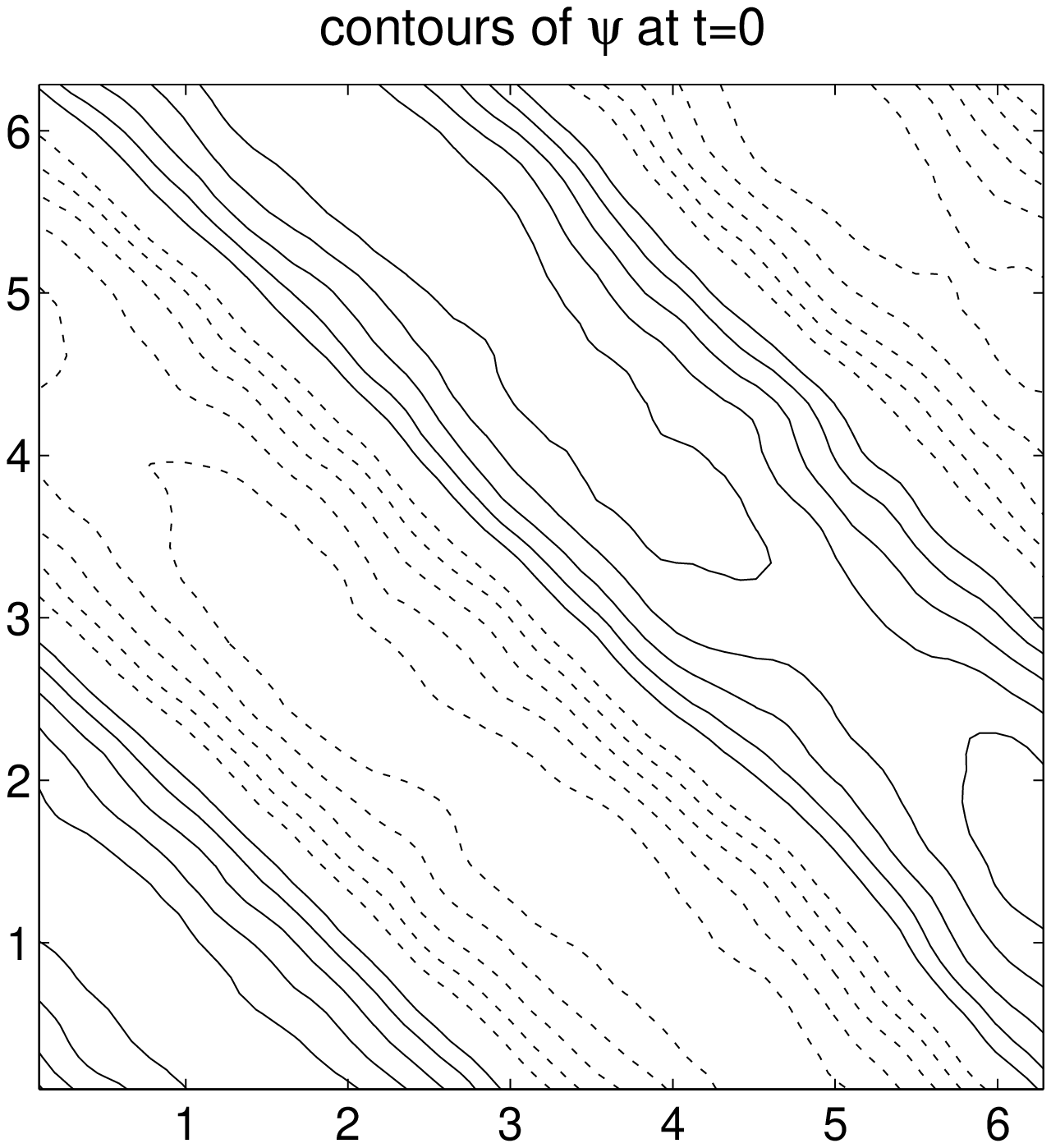}}
\end{minipage}
\begin{minipage}[c]{.34 \linewidth}
\scalebox{1}[1.]{\includegraphics[width=\linewidth]{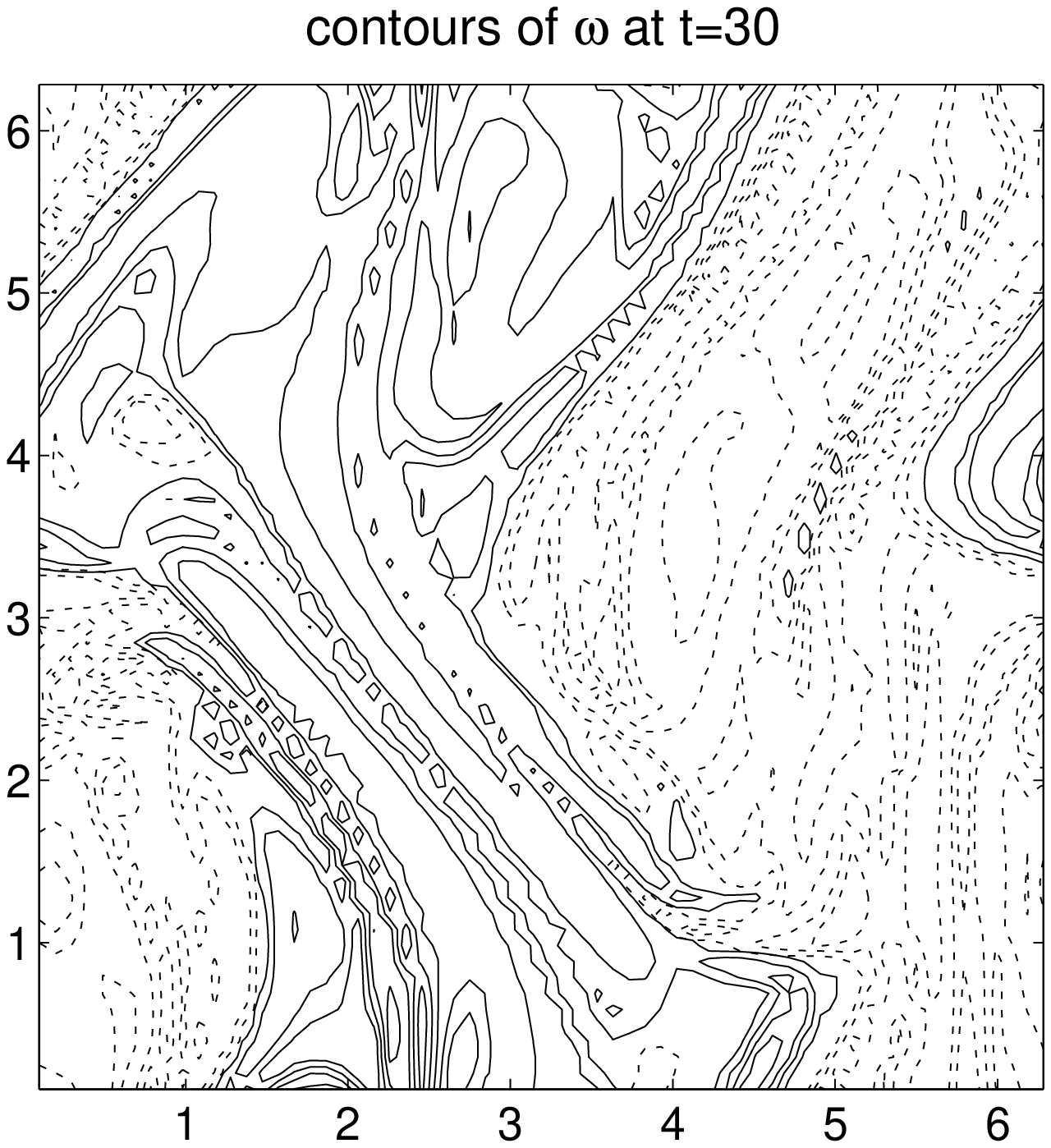}}
\end{minipage}
\hspace{0.5in}
\begin{minipage}[c]{.34 \linewidth}
\scalebox{1}[1.]{\includegraphics[width=\linewidth]{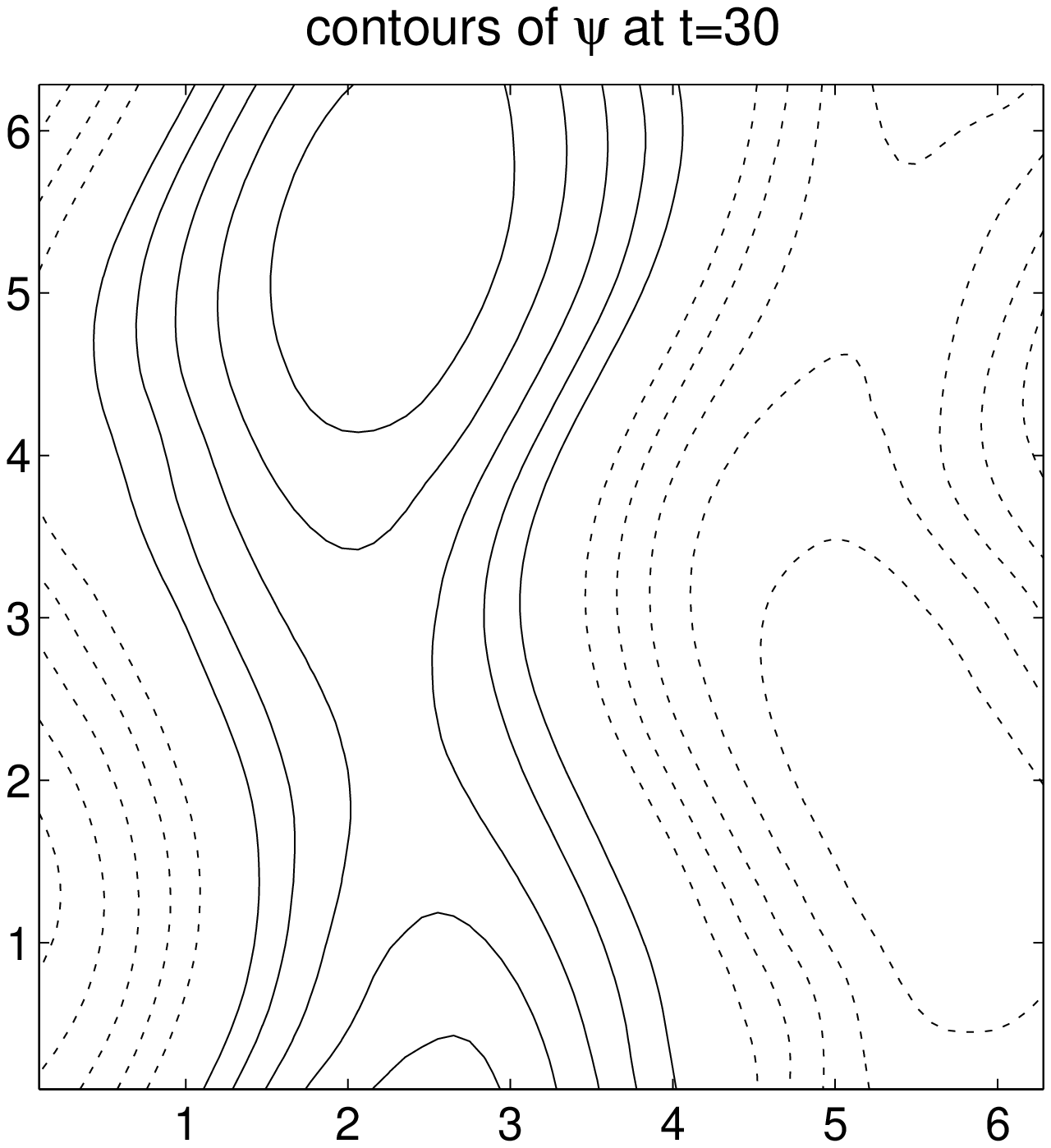}}
\end{minipage}
\begin{minipage}[c]{.34 \linewidth}
\scalebox{1}[1.]{\includegraphics[width=\linewidth]{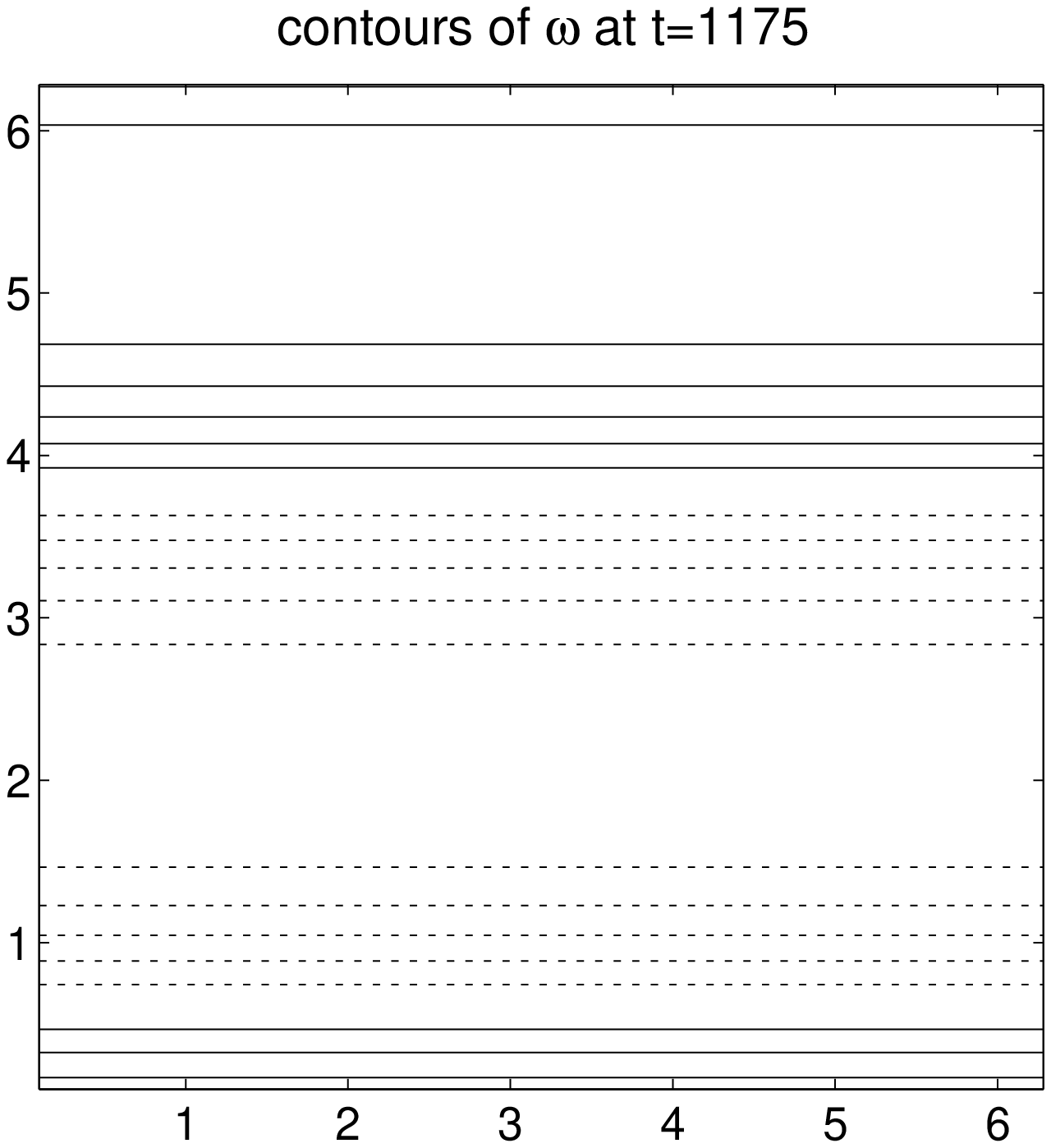}}
\end{minipage}
\hspace{0.5in}
\begin{minipage}[c]{.34 \linewidth}
\scalebox{1}[1.]{\includegraphics[width=\linewidth]{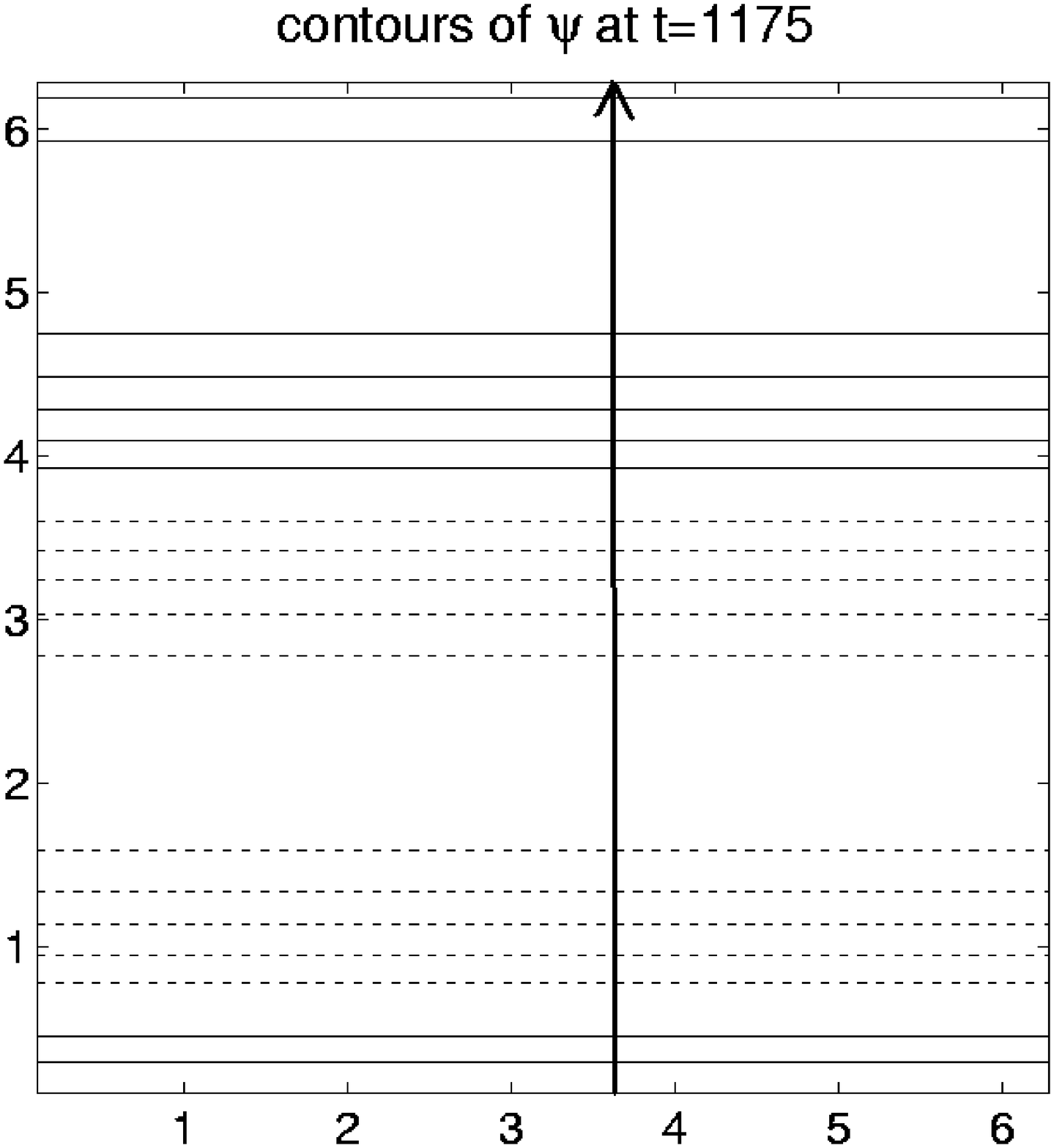}}
\end{minipage}
\caption{Contours of constant vorticity (left column) and constant
stream function (right column) at three different times for the
run with a new initial condition leading to the ``bar'' final
state. The arrow line in the last figure of the right column is
used to mark the flow field, and calculate the effective area  in
Fig. 8.} \label{fig:morebar3}
\end{figure*}

\begin{figure*}[!htbp]
\begin{minipage}[c]{.9 \linewidth}
\scalebox{1}[1]{\includegraphics[width=\linewidth]{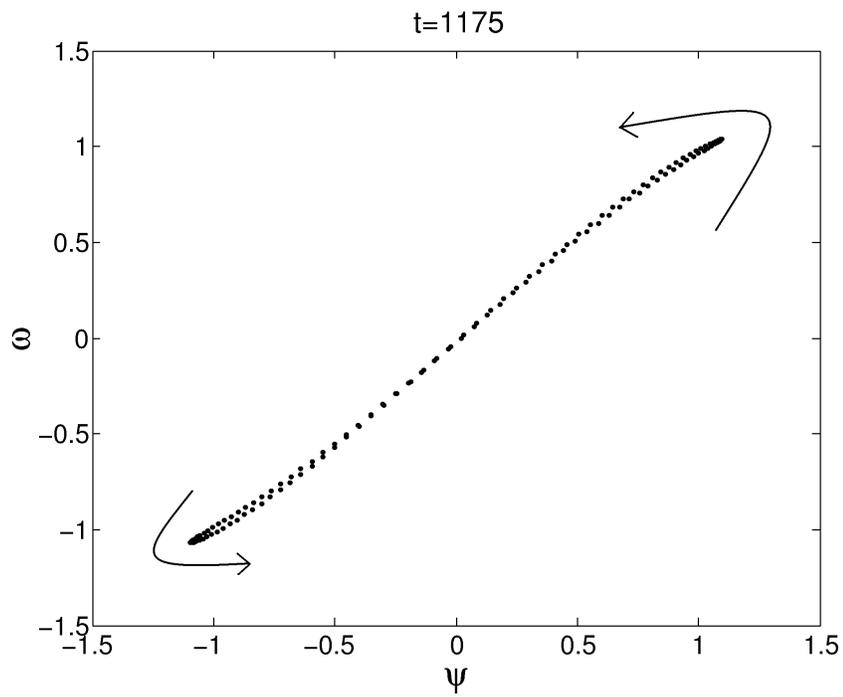}}
\end{minipage}
\caption{The $\omega-\psi$ scatter plot for the run shown in Figs.
7. The two ends of the plot are actually two loops. The two arrows
indicate the orientations of these two loops, which are obtained
by finding the corresponding points along the arrow line in the
last plot of Figs. 7.} \label{fig:morebar4}
\end{figure*}

\newpage

\begin{figure*}[!htbp]
\centering
\begin{minipage}[c]{.5635 \linewidth}
\scalebox{1}[.95]{\includegraphics[width=\linewidth]{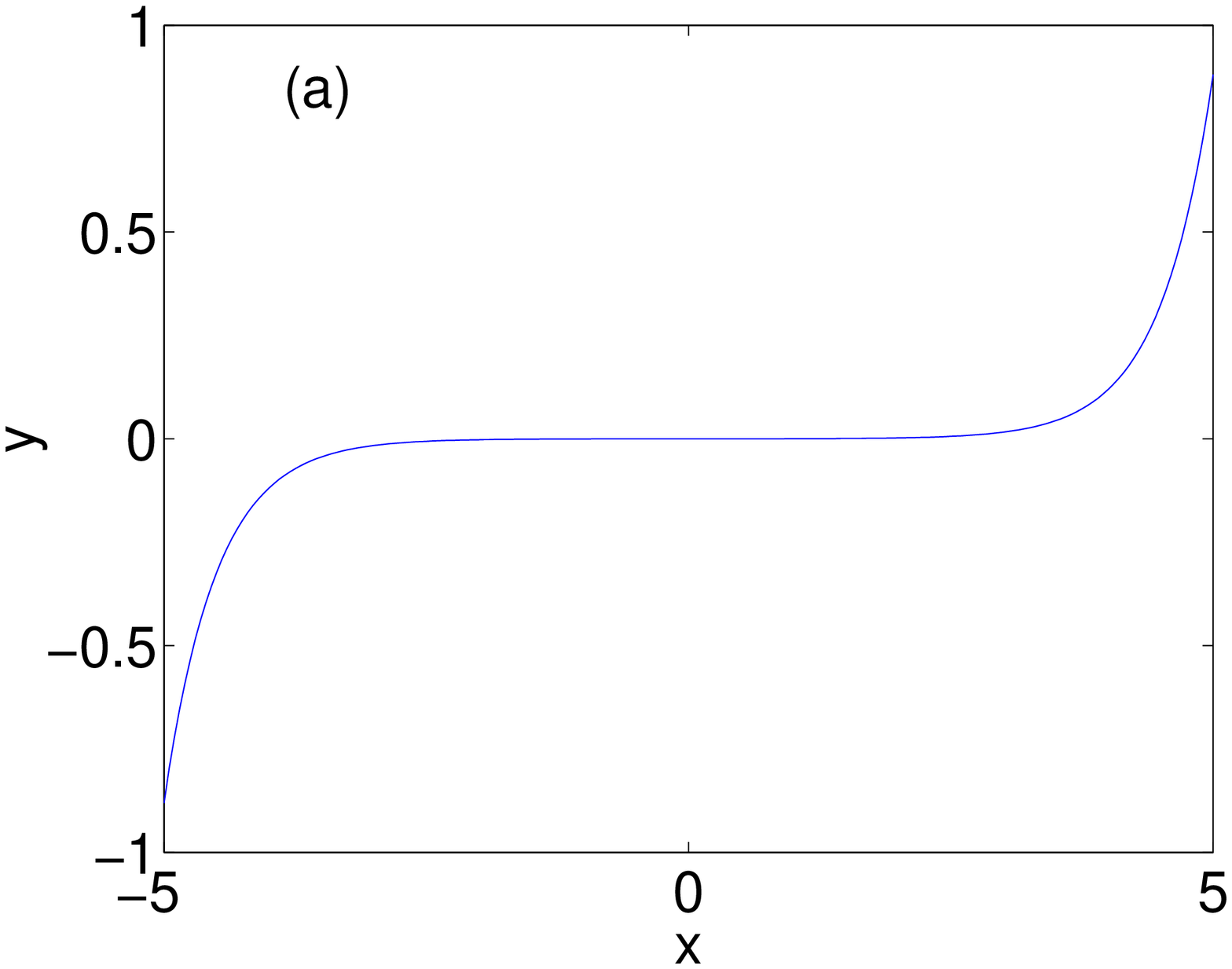}}
\end{minipage}
\begin{minipage}[c]{.54602 \linewidth}
\scalebox{1}[.95]{\includegraphics[width=\linewidth]{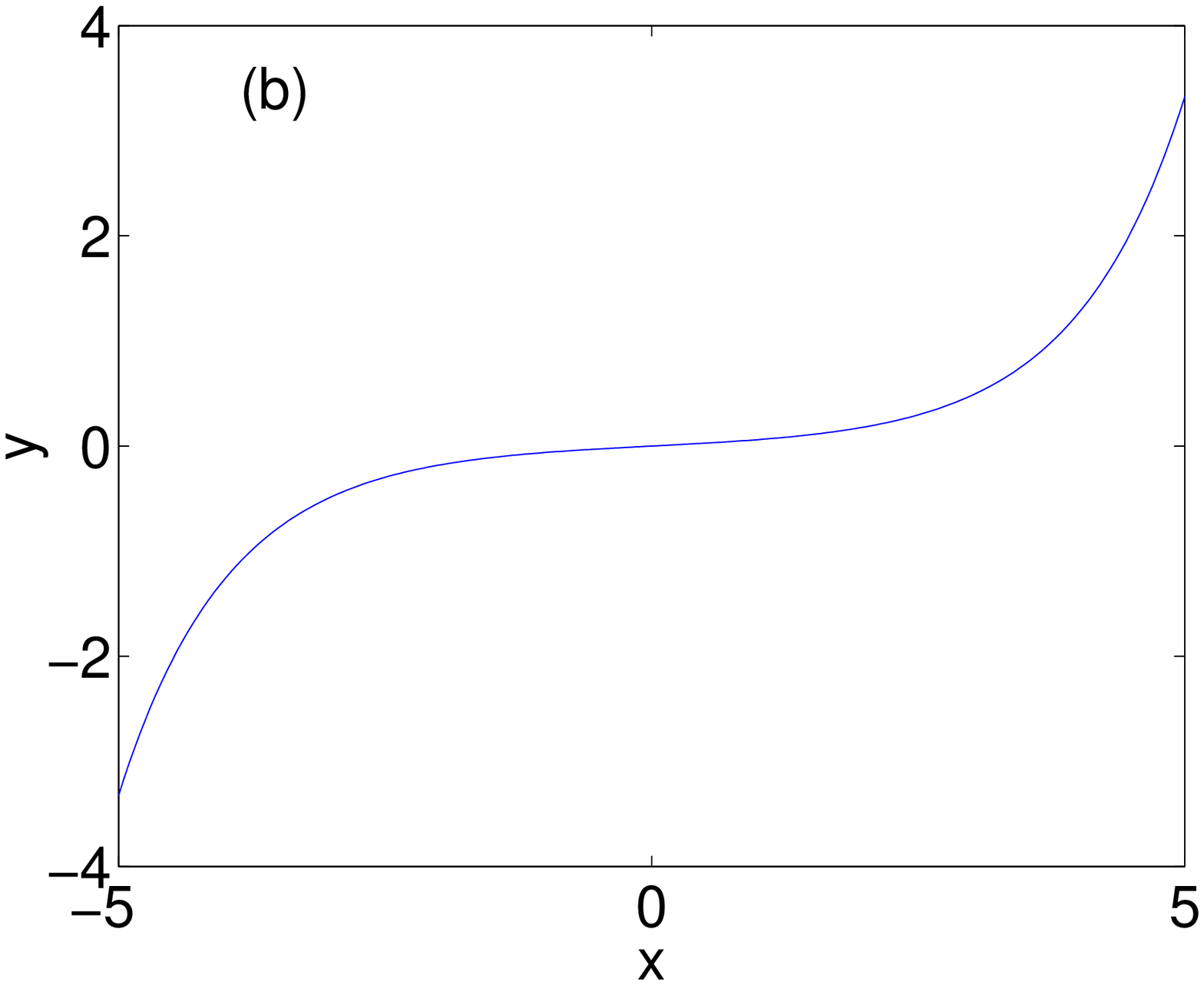}}
\end{minipage}
\begin{minipage}[c]{.5635 \linewidth}
\scalebox{1}[.95]{\includegraphics[width=\linewidth]{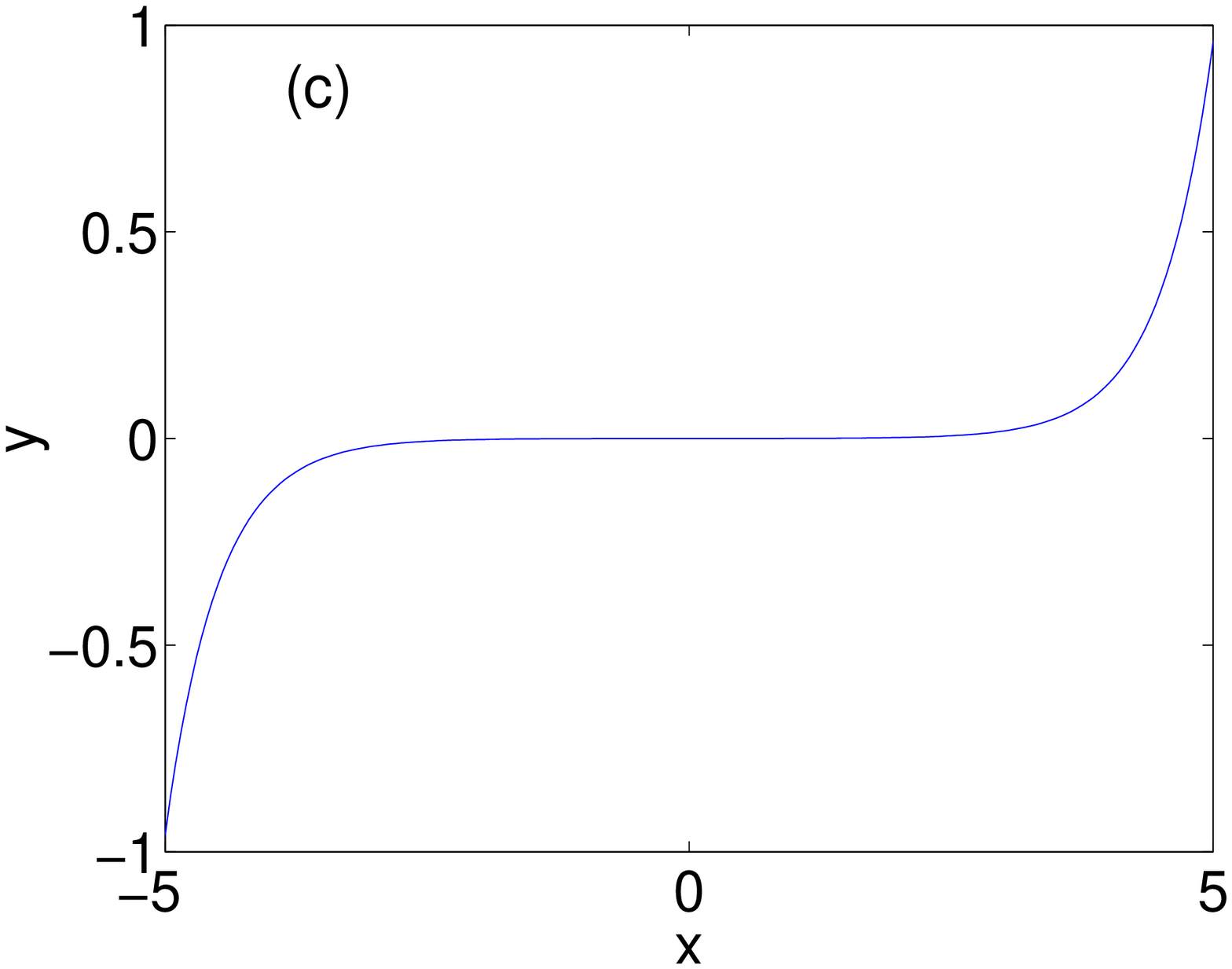}}
\end{minipage}
\caption{(a) is the plot of one simple sinh function (see Eq.
(11)). (b) and (c) are plots of two sums of 100 sinh functions
(see Eq. (12) and Eq. (13)); they all look like simple sinh
functions.} \label{fig:fit1}
\end{figure*}

\begin{figure*}[!htbp]
\centering
\begin{minipage}[c]{.68 \linewidth}
\scalebox{1}[1]{\includegraphics[width=\linewidth]{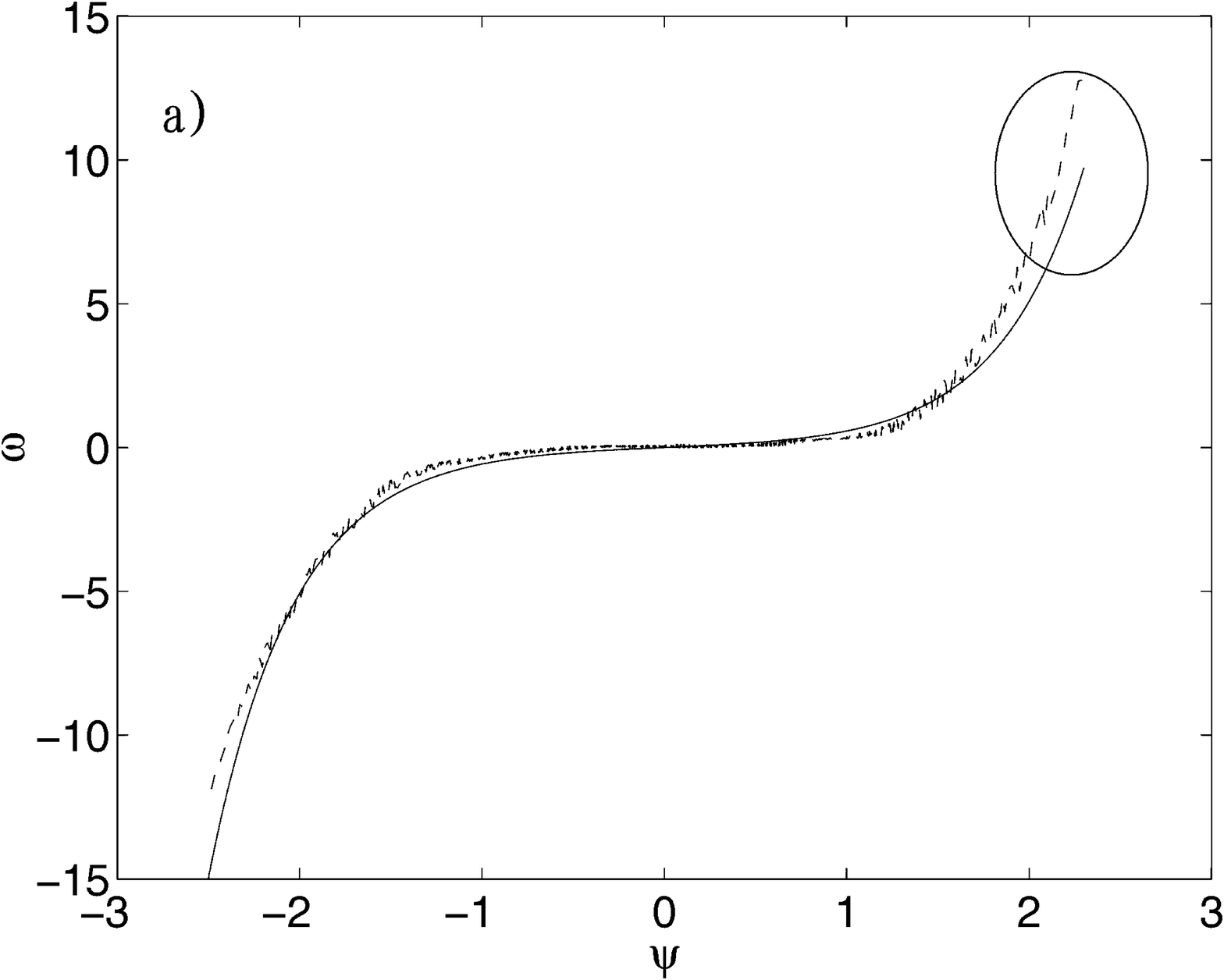}}
\end{minipage}
\begin{minipage}[c]{.68 \linewidth}
\scalebox{1}[1]{\includegraphics[width=\linewidth]{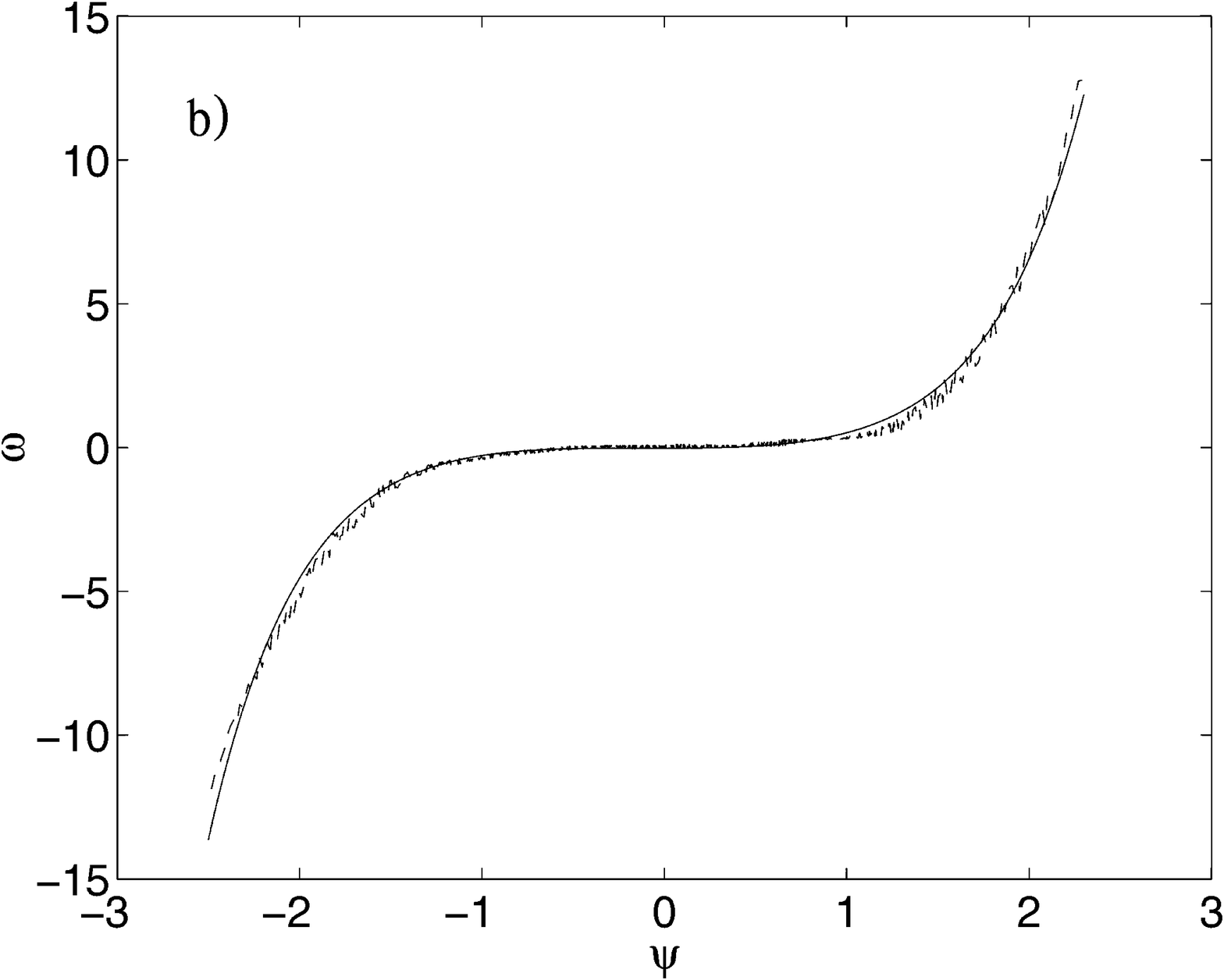}}
\end{minipage}
\caption{Different functions (indicated by curves drawn through
the plotted points) are used to fit the same $\omega-\psi$ scatter
plot of the ``dipole'' final state. (a) The fitting function is $y
= 0.13 sinh (2.16 x)$, the correlation factor is $R^2 = 0.95$. (b)
The fitting function is $y = 0.73 sinh(1.71 x) - 0.54 e^{1.07x} +
0.52 e^{- 1.27 x}$, the correlation factor is $R^2 = 0.99$.}
\label{fig:fit2}
\end{figure*}


\begin{thebibliography}{99}
\bibitem{kn:q2}
 Such relation, which is also called ``the multi-valued
function'' sometimes, is different from the ordinary definition of
the function because the derivatives of it can not be defined. In
Eq. (\ref{eq5}), the relation we use to derive $J(\omega ,\psi ) =
0$ is the normal sense of function, which is so-called one-to-one
or many-to-one function ($i.e.$ for any value of $x$, we can get
only one value of $y$ through relation $y = f(x))$.
\end{thebibliography}
\end{document}